\documentclass[10pt,a4paper,twoside,twocolumn,english,5p, authoryear]{elsarticle}
\usepackage{lmodern}

\usepackage[T1]{fontenc}
\usepackage[latin9]{inputenc}
\usepackage{color}
\usepackage{babel}
\usepackage{array}
\usepackage{textcomp}
\usepackage{url}
\usepackage{amsmath}
\usepackage{amssymb}
\usepackage{graphicx}
\usepackage[unicode=true,
 bookmarks=true,bookmarksnumbered=false,bookmarksopen=false,
 breaklinks=true,pdfborder={0 0 0},pdfborderstyle={},backref=false,colorlinks=true]
 {hyperref}
\hypersetup{
 pdfauthor={P. H. Hasselmann}}

\makeatletter

\pdfpageheight\paperheight
\pdfpagewidth\paperwidth

\newcommand{\lyxmathsym}[1]{\ifmmode\begingroup\def\b@ld{bold}
  \text{\ifx\math@version\b@ld\bfseries\fi#1}\endgroup\else#1\fi}

\providecommand{\tabularnewline}{\\}
\newcommand{\lyxdot}{.}

\@ifundefined{date}{}{\date{}}
\hypersetup{urlcolor=blue, citecolor=blue}
\usepackage{graphicx}
\usepackage{svg}

\@ifundefined{showcaptionsetup}{}{%
 \PassOptionsToPackage{caption=false}{subfig}}
\usepackage{subfig}
\AtBeginDocument{
  
}

\makeatother

\begin{document}

\begin{frontmatter}{}

\title{\textbf{Modeling optical roughness and first-order scattering processes
from OSIRIS-REx color images of the rough surface of asteroid (101955)
Bennu}}

\author[1]{Pedro H. Hasselmann}

\ead{pedro.hasselmann@obspm.fr}

\author{Sonia Fornasier$^{1,2}$}

\author{Maria A. Barucci$^{1}$}

\author{Alice Praet$^{1}$}

\author{Beth E. Clark$^{3}$}

\author{Jian-Yang Li$^{4}$}

\author{Dathon R. Golish$^{5}$}

\author{Daniella N. DellaGiustina$^{5}$}

\author{Jasinghege Don P. Deshapriya$^{1}$}

\author{Xian-Duan Zou$^{4}$}

\author{Mike G. Daly$^{6}$}

\author{Olivier S. Barnouin$^{7}$}

\author{Amy A. Simon$^{8}$}

\author{Dante S. Lauretta$^{5}$}

\address{\noindent \begin{flushleft}
\emph{1. LESIA, Observatoire de Paris, PSL, CNRS, Universit� de Paris,
Sorbonne Universit�, 5 place Jules Janssen, Meudon, France. }\\
\emph{2. Institut Universitaire de France (IUF), 1 rue Descartes,
Paris, France.}\\
\emph{3. Department of Physics and Astronomy, Ithaca College, Ithaca,
NY, USA.}\\
\emph{4. Planetary Science Institute, Tucson, AZ, USA.}\\
\emph{5. Lunar and Planetary Laboratory, University of Arizona, Tucson,
AZ, USA.}\\
\emph{6. The Centre for Research in Earth and Space Science, York
University, Toronto, Ontario, Canada.}\\
\emph{8. NASA Goddard Space Flight Center, Greenbelt, MD, USA}\\
\emph{7. The Johns Hopkins University Applied Physics Laboratory,
Laurel, MD, USA.}
\par\end{flushleft}}
\begin{abstract}
The dark asteroid (101955) Bennu studied by NASA\textquoteright s
OSIRIS-REx mission has a boulder-rich and apparently dust-poor surface,
providing a natural laboratory to investigate the role of single-scattering
processes in rough particulate media. Our goal is to define optical
roughness and other scattering parameters that may be useful for the
laboratory preparation of sample analogs, interpretation of imaging
data, and analysis of the sample that will be returned to Earth. We
rely on a semi-numerical statistical model aided by digital terrain
model (DTM) shadow ray-tracing to obtain scattering parameters at
the smallest surface element allowed by the DTM (facets of \textasciitilde{}10
cm). Using a Markov Chain Monte Carlo technique, we solved the inversion
problem on all four-band images of the OSIRIS-REx mission\textquoteright s
top four candidate sample sites, for which high-precision laser altimetry
DTMs are available. We reconstructed the \emph{a posteriori} probability
distribution for each parameter and distinguished primary and secondary
solutions. Through the photometric image correction, we found that
a mixing of low and average roughness slope best describes Bennu's
surface for up to $90^{\circ}$ phase angle. We detected a low non-zero
specular ratio, perhaps indicating exposed sub-centimeter mono-crystalline
inclusions on the surface. We report an average roughness RMS slope
of $27_{-5}^{\circ+1}$, a specular ratio of $2.6_{-0.8}^{+0.1}\%$,
an approx. single-scattering albedo of $4.64_{-0.09}^{+0.08}\%$ at
550 nm, and two solutions for the back-scatter asymmetric factor,
$\xi^{(1)}=-0.360\pm0.030$ and $\xi^{(2)}=-0.444\pm0.020$, for all
four sites altogether.
\end{abstract}

\end{frontmatter}{}

\section{{\small{}Introduction}}

{\footnotesize{}OSIRIS-REx (Origins, Spectral Interpretation, Resource
Identification, and Security\textendash Regolith Explorer) is a NASA
mission intended to collect and bring back to Earth a sample of pristine
material from the carbonaceous chondrite\textendash like asteroid
(101955) Bennu \citep{2017SSRv..212..925L}. Arriving at Bennu on
December 3, 2018, the mission has performed disk-resolved surface
characterization to better understand the asteroid and prepare for
the selection of a sample site. The spacecraft\textquoteright s remote
sensing payload includes a VIS camera suite (OCAMS), a scanning laser
altimeter (OLA), two point spectrometers (OVIRS and OTES; VIS-NIR
and thermal IR, respectively) and an x-ray imaging spectrometer (REXIS). }{\footnotesize \par}

{\footnotesize{}The initial results from the mission confirmed the
presence of an equatorial budge \citep{2019NatAs...3..352S,2020P&SS..18004764B}
and aqueously altered minerals with similar compositions to those
found in CM carbonaceous chondrites \citep{2019NatAs...3..332H}.
The OCAMS images showed a dark, boulder-rich environment with an average
geometric albedo of $4.4\pm0.2\%$ at 550 nm. Multiple instruments
indicated a lack of widespread micrometric grains \citep{2019NatAs...3..341D,2019Natur.568...55L}.}{\footnotesize \par}

{\footnotesize{}In this work, we study the role of multi-scale roughness,
shadowing, and other first-order scattering processes on the surface
of Bennu. This asteroid\textquoteright s dark, boulder-rich, apparently
dust-poor surface provides a natural laboratory to investigate the
role of single-scattering processes in rough particulate surfaces
and their effects on the bi-directional reflectance distribution function
(BRDF) or radiance factor (RADF) distribution. }{\footnotesize \par}

{\footnotesize{}For highly absorbent surfaces observed off the opposition
configuration, the RADF distribution is largely controlled by roughness
with a characteristic scale much larger than the particle size, i.e.,
the roughness scale situated in the optical regime. This regime configuration
is also known as hierarchically arranged random topography \citep{2005Icar..173....3S}.
On rough surfaces, there are the formation of shadows and occlusions,
yielding most of the variation in reflectance of an homogeneous surface
observed at varied scattering geometry. }{\footnotesize \par}

{\footnotesize{}For analytically computing the radiative contribution
of the macroscopic roughness, the Hapke shadowing function \citep{1984Icar...59...41H}
has been usually adopted by the planetary science community \citep{2015aste.book..129L}.
However, this function has come under scrutiny for failing to reproduce
non-Gaussian topographies \citep{2015Icar..252....1D,2017Icar..290...63L},
poorly scaling for higher roughness slopes \citep{2017Icar..290...63L}
and allegedly violating the energy conservation \citep{2012JQSRT.113.2431S,2013JQSRT.116..184H}.
To counterpoint these three problems from the Hapke shadowing function,
we reintroduce the formalism put forward by \citet{1998ApOpt..37..130V},
a semi-numerical statistical model that simulates diffuse and specular
scattering arising from illuminated Gaussian-random rough surfaces
that scales into high roughness slopes. On its first application to
astronomical data, \citet{2010Icar..208..548G} adapted the model
to use the Lommel-Seeliger law, and it was successfully applied to
ROLO (Robotic Lunar Observatory) photometric data of the Moon. The
results showed generally good agreement with the Hapke model, but
with a more pronounced optical roughness for the Lunar Highlands \citep{1999Icar..141..107H}.
The model has some advantages over the Hapke shadowing function: its
formalism can accomodate any scattering law \citep{1941ApJ....93..403M,1976SvA...19...385388,2005JRASC..99...92F},
any statistical continuous slope distributions, and also takes into
account inter-reflection. The model remains mathematically fairly
simple and can be also applied to photometrically correct images and
spectra \citep{2011P&SS...59.1326S}.}{\footnotesize \par}

{\footnotesize{}Tackling the surface roughness slope is also limited
by the spatial resolution of data and the shadow effects of meso-scale
topography such as boulder fields. \citet{2005Icar..173....3S} has
demonstrated that scattering ``boulder-like'' features over the
soil can significantly change the shadowing function for intermediary
phase angles. As the OSIRIS-REx mission has the capability to generate
accurate digital terrain models (DTMs) from data acquired by OLA \citep{2017SSRv..212..899D,2020P&SS..18004764B},
we can directly ray-trace the sub-pixel shadowing using the provided
DTMs. Ray-tracing techniques have been widely used by the photometric
astronomical community to theoretically check the validity of photometric
models, but seldom applied to the direct photometric correction of
remote sensing data.}{\footnotesize \par}

{\footnotesize{}In our study of Bennu, we model optical roughness
and first-scattering processes following \citet{1998ApOpt..37..130V}
and using the four-band color images obtained by the OCAMS MapCam
imager \citep{2018SSRv..214...26R,Golish2020}. Our goal is to reintroduce
a consistent framework where rough surfaces can be mathematically
treated without losing effectiveness to provide a photometric correction.
Photometric data correction is a fundamental product for spatially
resolved data, and it is required for the inter-comparison of data
obtained under different observational conditions and the albedo standardization
of all data. Furthermore, by relying on direct numerical modeling,
we can obtain a precise estimate of the surface roughness slope that
will support laboratory preparations of surface analogs and interpretation
of the micro-physics of the returned sample.}{\footnotesize \par}

{\footnotesize{}We also introduce a new tool for rendering DTMs into
varied instrumental fields of view (FOVs), ray-tracing shadows at
sub-pixel accuracy, and obtaining the necessary geometric and solid
angles per DTM surface element. The inverse problem is solved using
the Markov Chain Monte Chain (MCMC) technique to obtain}\emph{\footnotesize{}
a posteriori}{\footnotesize{} probability distributions of the model
parameters. MCMC was chosen for its capability to describe non-unique
solutions and deal with heteroscedasticity within the sample and the
model \citep{2015Icar..260...73S}.}{\footnotesize \par}

\section{{\small{}OSIRIS-REx MapCam images of sample site candidates}}

{\scriptsize{}}
\begin{table*}[t]
{\scriptsize{}\caption{{\footnotesize{}\label{tab:data_specs}OCAMS images obtained during
the Equatorial Stations campaign \citep{2020Icarus...GOLISH}. S/C
for spacecraft.}}
}{\scriptsize \par}
\centering{}%
\begin{tabular}{ccccccc}
\hline 
{\footnotesize{}Station} & {\footnotesize{}Date (YYYY-MM-DD)} & {\footnotesize{}$N$ color images} & {\footnotesize{}S/C Distance (km)} & {\footnotesize{}meter/pixel} & {\footnotesize{}phase angle range} & {\footnotesize{}Local Time}\tabularnewline
\hline 
\hline 
{\footnotesize{}EQ1} & {\footnotesize{}2019-04-25} & {\footnotesize{}550} & {\footnotesize{}4.97-5.09} & {\footnotesize{}$0.32-0.34$} & {\footnotesize{}$43^{\circ}\lyxmathsym{\textendash}47^{\circ}$} & {\footnotesize{}3:00 pm}\tabularnewline
\hline 
{\footnotesize{}EQ2} & {\footnotesize{}2019-05-02} & {\footnotesize{}387} & {\footnotesize{}4.87-4.98} & {\footnotesize{}``} & {\footnotesize{}$130^{\circ}\lyxmathsym{\textendash}134^{\circ}$} & {\footnotesize{}3:20 am}\tabularnewline
\hline 
{\footnotesize{}EQ3} & {\footnotesize{}2019-05-09} & {\footnotesize{}550} & {\footnotesize{}4.85-4.99} & {\footnotesize{}``} & {\footnotesize{}$7^{\circ}\lyxmathsym{\textendash}11^{\circ}$} & {\footnotesize{}12:30 pm}\tabularnewline
\hline 
{\footnotesize{}EQ4} & {\footnotesize{}2019-05-16} & {\footnotesize{}545} & {\footnotesize{}4.80-4.95} & {\footnotesize{}``} & {\footnotesize{}$28^{\circ}\lyxmathsym{\textendash}32^{\circ}$} & {\footnotesize{}10:00 am}\tabularnewline
\hline 
{\footnotesize{}EQ5} & {\footnotesize{}2019-05-23} & {\footnotesize{}690} & {\footnotesize{}4.84-4.95} & {\footnotesize{}``} & {\footnotesize{}$89^{\circ}\lyxmathsym{\textendash}93^{\circ}$} & {\footnotesize{}6:00 am}\tabularnewline
\hline 
{\footnotesize{}EQ6} & {\footnotesize{}2019-05-30} & {\footnotesize{}500} & {\footnotesize{}4.99-5.17} & {\footnotesize{}``} & {\footnotesize{}$130^{\circ}\lyxmathsym{\textendash}134^{\circ}$} & {\footnotesize{}8:40 pm}\tabularnewline
\hline 
{\footnotesize{}EQ7} & {\footnotesize{}2019-06-06} & {\footnotesize{}555} & {\footnotesize{}4.93-5.05} & {\footnotesize{}``} & {\footnotesize{}$89^{\circ}\lyxmathsym{\textendash}93^{\circ}$} & {\footnotesize{}6:00 pm}\tabularnewline
\hline 
\end{tabular}
\end{table*}
{\footnotesize{}MapCam is equipped with four band color filters (60-90
nm wide) centered at 473 ($b^{\prime}$), 550 (v), 698 (w), and 847
(x) nm, in the visible range. The images are projected in 1024x1024
pixel CCD with a FOV of $4^{\circ}\times4^{\circ}$ \citep{2018SSRv..214...26R}.
The images are radiometrically calibrated into RADF and corrected
for any optical distortion \citep{Golish2020}. }{\footnotesize \par}

{\footnotesize{}The photometric data analyzed in this work were acquired
during the Equatorial Stations campaign (EQ), a subphase of the Detailed
Survey mission phase in spring 2019. MapCam acquired 3,784 multi-band
images over a full rotation per station of (101955) Bennu (4.3 hours)
at a distance of about 5 km. The spacecraft was approximately placed
over the asteroid's equator, reached after a series of polar hyperbolic
trajectories. At this distance, the spatial pixel resolution at nadir
subtended about 33 cm of Bennu's surface. EQ included seven observational
configurations at different local solar times of day, imaging the
asteroid at five different phase angles, $\alpha=[7.5\text{\textdegree},30\text{\textdegree},45\text{\textdegree},90\text{\textdegree},130\text{\textdegree}]$.
No data during opposition were obtained in this campaign, so our analysis
does not include any modeling with respect to the opposition effect.
Table \ref{tab:data_specs} summarizes the information for each EQ.}{\footnotesize \par}

{\footnotesize{}High-precision DTMs have proved important to obtaining
precise geometric angles and can heavily affect photometric corrections
\citep{2020Icarus...GOLISH}. Here we analyzed the pixels subtended
by high-precision DTMs (10 cm ground sample distance) of the OSIRIS-REx
mission\textquoteright s top four candidate sample sites; these DTMs
were generated from OLA scans performed during the Orbital B mission
phase in summer 2019 \citep{2017SSRv..212..899D,2020P&SS..18004764B}.
The candidate sample sites were selected by the mission following
criteria for safety, sampleability, deliverability, and scientific
value. These four primary candidates were called Sandpiper ($latitude=-47^{\circ}$,
$longitude=322^{\circ}$), Osprey ($11^{\circ},$$88^{\circ}$), Nightingale
($56^{\circ},$$43^{\circ}$), and Kingfisher ($11^{\circ},$$56^{\circ}$).
The varied latitudes and longitudes of the sites provides the range
of observational conditions required for our analysis. The DTM zones
are a square of 50 m scanline length, about two times the length of
the actual sample sites therein. They have a flat surface of approx.
2500 m\textsuperscript{{\footnotesize{}2}}.}{\footnotesize \par}

\section{{\small{}Shapeimager: Scattering geometry \& macro-shadows}}

{\footnotesize{}To study the precise dependence of the RADF on the
incidence (}\emph{\footnotesize{}i}{\footnotesize{}), emergence (}\emph{\footnotesize{}e}{\footnotesize{}),
azimuth ($\varphi$) and phase ($\alpha$) angles \citep{2011P&SS...59.1326S},
we need these angles at sub-pixel resolution. Also called scattering
geometry conditions, (}\emph{\footnotesize{}i}{\footnotesize{}, }\emph{\footnotesize{}e}{\footnotesize{},
$\varphi$,$\alpha$) are obtained through FOV renderings. The renderings
depend on DTM, the target and the observer solar and relative positions,
as well as the detector optical specifications. The smallest rendered
surface elements are the triangular facets of the DTM. For this work,
we used the 10-cm OLA DTMs and reconstructed ephemeris and detector
specifications using NAIF SPICE kernels \citep{1996P&SS...44...65A,2018P&SS..150....9A}
provided by the OSIRIS-REx Flight Dynamics System. To obtain a precise
representation of a surface under a detector, we must also incorporate
the scattering surface properties, as well as shadows. For this task,
we developed a set of Python tools}\footnote{{\footnotesize{}available at }\url{https://github.com/pedrohasselmann/shapeimager}}{\footnotesize{}
for disk-resolved FOV \& image renderings called Shapeimager \citep{2019EPSC...13..225H}.
The purpose of Shapeimager is to obtain the most precise geometric
information for Solar System objects observed by any mission-detector
configuration. Its crucial feature is the facet-scale calculation
of macro-shadows out of any given shape model of any spatial precision.}{\footnotesize \par}

{\footnotesize{}Macro-shadows are computed at the sub-pixel level
if the images have smaller spatial resolution than the DTM. The image
plane is partitioned in the facet-scale or smaller, a pixeled image
is reproduced from the light source's point of view, and another is
produced from the observer's point of view. For each partition/pixel,
two rays are traced: one from the light source and another from the
observer. If the ray is intercepted in any of the two cases, we have
a facet that is shadowed, occluded, or both.}{\footnotesize \par}

{\footnotesize{}Shapeimager tracks the position, orientation, solid
angle, and all necessary geometric angles of every visible facet.
If available, the instrumental point-spread function (PSF) can also
be taken into account when computing the intensity contribution of
each single facet into the total flux. For the full mathematical framework
behind image rendering, we recommend readers see \citet{Hartley2004}.}{\footnotesize \par}

\begin{figure}
\begin{centering}
{\footnotesize{}\includegraphics[scale=0.62]{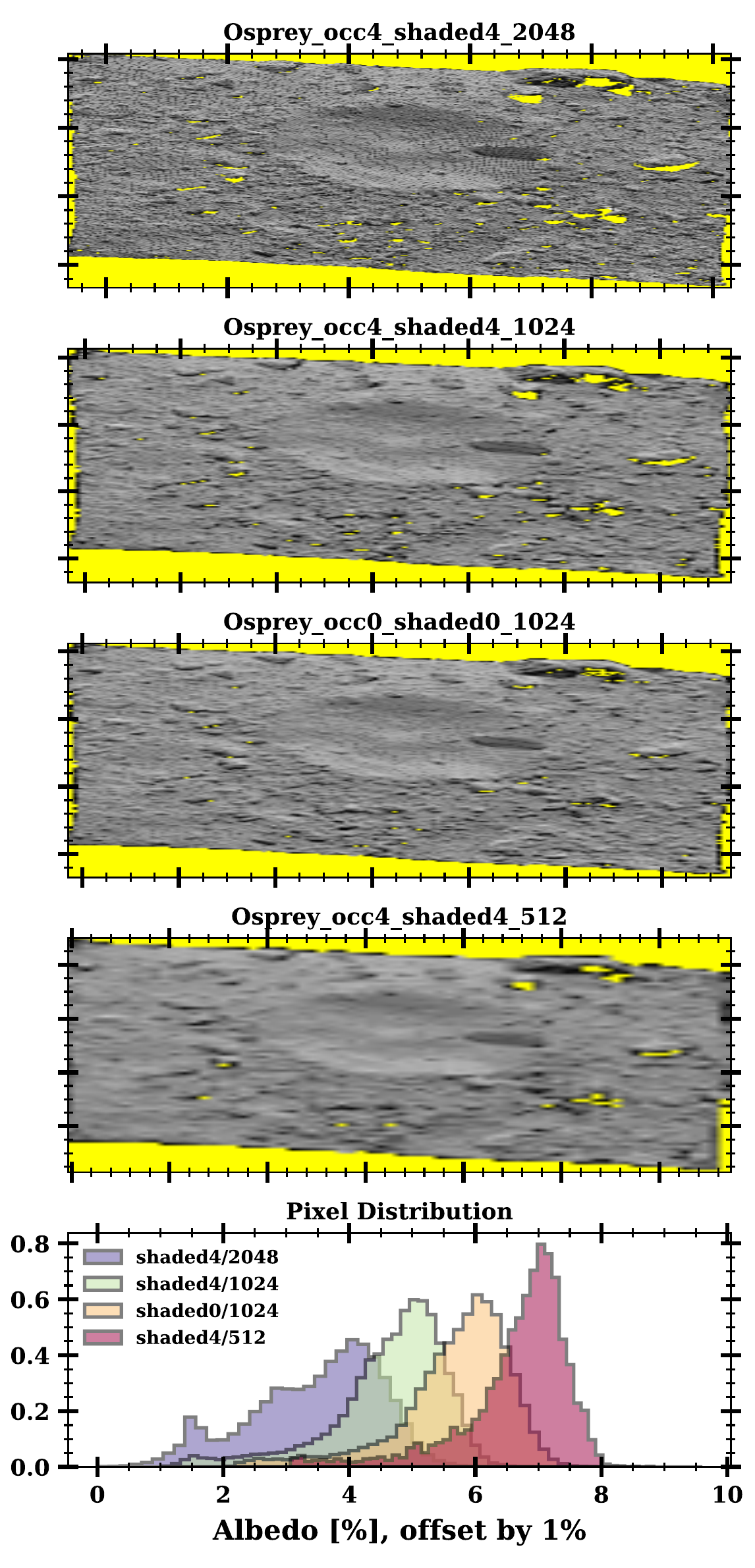}}
\par\end{centering}{\footnotesize \par}
{\footnotesize{}\caption{\label{fig:Example-of-rendering}{\footnotesize{}Shadow ray-tracing
of Osprey OLA DTM site in the MapCam instrument settings at UTC 2019-04-25,
18:04:04.000 ($\alpha=44^{\circ}$). The brightness profile is calculated
using the Lommel-Seeliger law multiplied by Bennu's geometric albedo
\citep{2019NatAs...3..341D}. All images are constrained to same contrast
and brightness levels. Null values are color-coded in yellow. From
the first to the fourth panel: 2048x2048 (shadowed), 1024x1024 (shadowed,
Standard OCAMS renderization), 1024x1024 (No shadows), 512x512 (shadowed).
The fifth panel shows the normalized distribution of albedo {[}\%{]}
per pixel for every case above, offset by 1\%.}}
}{\footnotesize \par}
\end{figure}

\begin{figure}
\begin{centering}
\includegraphics[scale=0.45]{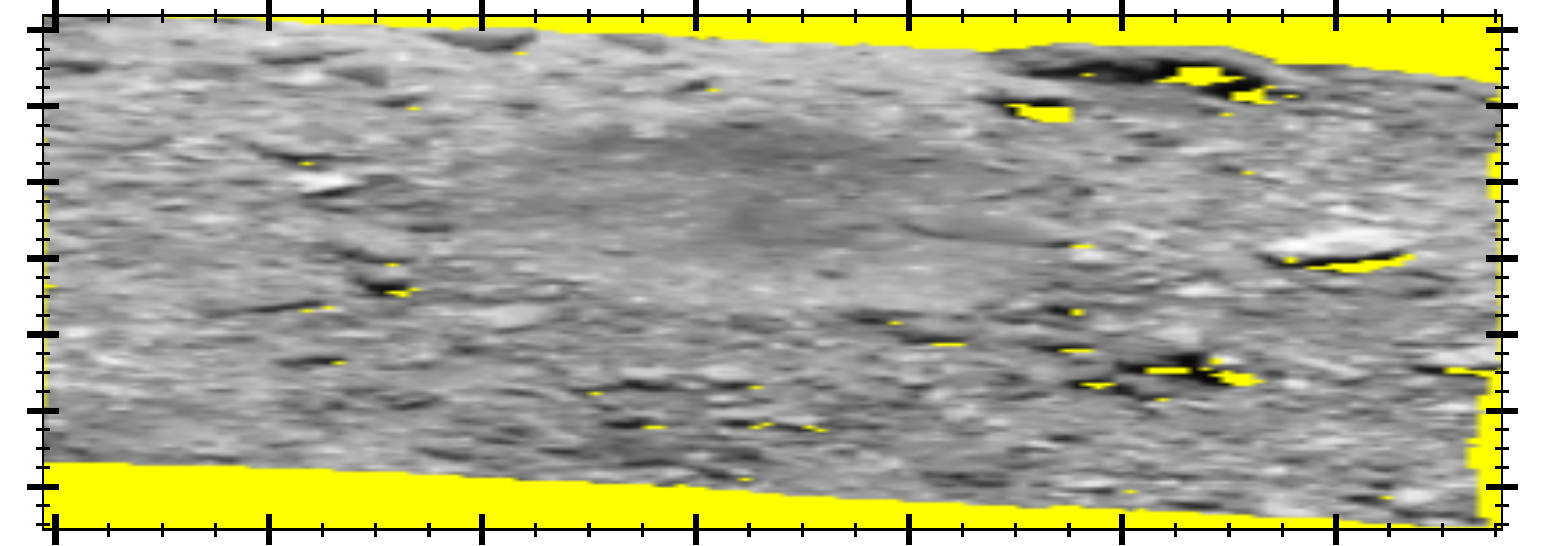}
\par\end{centering}
{\footnotesize{}\caption{{\footnotesize{}\label{fig:Osprey_example}Example of a MapCam x-filter
image segment of Osprey. Image taken at UTC 2019-04-25, 18:04:04.000
($\alpha=44^{\circ}$). Values smaller than 5e-5 are color-coded in
yellow.}}
}{\footnotesize \par}

\end{figure}

{\footnotesize{}To exemplify the results that can be obtained with
the Shapeimager, we present an example for four different CCD pixel
size and shadow tracing (Figure \ref{fig:Example-of-rendering}) of
rendering of the OLA DTM of the Osprey site in the MapCam FOV at UTC
April 25 2019, 18:04:04 ($\alpha=44^{\circ}$, Figure \ref{fig:Osprey_example}).
From the first to the fourth panel we can perceive that small shadow
and shading structures become gradually absorbed into the pixel size
as we increase the CCD grid by 4 times. For the third panel, we have
no shadowing, only shading due to the Lommel-Seeliger law. Resolving
shadows leads to effects in the brightness distribution. The brightness
distribution becomes progressively less ``peaked''', and by the
$2048\times2048$ pixel grid resolution, which corresponds approximately
to 1 facet per pixel, a second peak is revealed around 1.7\% albedo
due to the DTM shading.}{\footnotesize \par}

{\footnotesize{}Therefore, to account for this ``instrumental effect'''
\textemdash{} i.e., the way in which surfaces are perceived through
varying pixel resolutions \textemdash{} we apply a sub-pixel shadow
and shading operation (Appendix A) when analyzing the photometrically
corrected images using the scattering roughness model (Section 4).
This operation allows us to reduce the effects of the boulder-field
topography. However, we remain limited by the DTM spatial resolution,
especially with respect to the pebble field of objects a few centimeters
in size, which is not captured in the 10-cm OLA DTMs that we used
and may influence the final root-mean-square (RMS) roughness slope.}
{\footnotesize{}}
{\footnotesize \par}

\section{{\small{}Scattering roughness model}}

\subsection{{\small{}Semi-numerical roughness model}}

{\footnotesize{}Multi-scale roughness comprises a major part of reflectance
variation observed on planetary and atmosphere-less small body surfaces
\citep{1999Icar..141..107H,2005Icar..173....3S,2017Icar..289..281C}.
Random ``tilts'' due to macroscopic and microscopic surface irregularities
can contribute more or less to the radiance distribution in certain
observational conditions regarding the incidence light. These ``tilts''
can mutually occlude or shadow themselves, giving rise to much photometric
variation.}{\footnotesize \par}

{\footnotesize{}The \citealp{1998ApOpt..37..130V} semi-numerical
roughness model that we use here is an alternative to the Hapke shadowing
function \citep{1984Icar...59...41H}. It assumes a scaling surface
with a Gaussian distribution of tilts in the geometrical optics regime.
The standard deviation and autocorrelation function determine the
roughness as RMS slope. The model goes further in considering the
number of tilts occluded and shadowed, to finally produce a set of
numerical-analytical equations describing the radiance out of any
given diffuse scattering law. We advise readers to see \citet{1998ApOpt..37..130V}
for the detailed mathematical framework. Here we summarize the relevant
equations, while keeping consistency of notations with \citet{2010Icar..208..548G},
the most recent description of the model.}{\footnotesize \par}

{\footnotesize{}Given a ``much-larger-than-wavelength'' particulate,
rough, and isotropic surface $dA$ where the normal vector $\hat{n}$
coincides with the }\emph{\footnotesize{}z}{\footnotesize{} axis,
the radiation is incident at an angle }\emph{\footnotesize{}i}{\footnotesize{}
(incidence unit vector $\hat{i}$) and observed at an angle }\emph{\footnotesize{}e}{\footnotesize{}
(emergence unit vector $\hat{e}$) relative to the same }\emph{\footnotesize{}z}{\footnotesize{}
axis. Azimuth $\varphi$ is an angle between $\hat{i}$ and $\hat{e}$
at the }\emph{\footnotesize{}xy}{\footnotesize{} plane orthogonal
to }\emph{\footnotesize{}z}{\footnotesize{} (see Fig. 1 in \citealp{1998ApOpt..37..130V}).
Phase angle $\alpha$ is another geometric angle between $\hat{i}$
and $\hat{e}$ but measured at the plane formed by these two unit
vectors instead. Considering that the rough surface $dA$ is characterized
by smaller local surfaces tilted (just ``tilt'' hereafter) at an
angle $\theta_{a}$ and azimuth $\varphi_{a}$ is normally distributed,
the probability distribution of tilts is:}{\footnotesize \par}

{\footnotesize{}
\begin{equation}
P_{a}(\theta_{a},\sigma)d\theta_{a}=\frac{\sin\theta_{a}}{\sigma^{2}\cos^{3}\theta_{a}}\exp\left(\frac{-\tan^{2}\theta_{a}}{2\sigma^{2}}\right)d\theta_{a}
\end{equation}
}{\footnotesize \par}

{\footnotesize{}The roughness is therefore characterized by a single
parameter, the RMS slope $\sigma$. This same Gaussian distribution
framework leads to the derivation of simplified equations for the
probability of a certain tilt to be both illuminated and visible:}{\footnotesize \par}

{\footnotesize{}
\begin{equation}
P_{ill+vis}(i,e,\varphi,\sigma)\approx1/\left\{ 1+\Lambda(\sigma,\max[i,e])+\xi\Lambda(\sigma,\min[i,e])\right\} 
\end{equation}
}{\footnotesize \par}
\noindent \begin{flushleft}
{\footnotesize{}Setting $\xi=4.41\varphi/(4.41\varphi+1)$ yields
an error never exceeding 3\% for $0<\sigma<1$ (From Eqs. 22, 23,
and 24 in \citealp{1998ApOpt..37..130V}). And for $\Lambda$ we have
\citep{1967ITAP...15..668S}:}
\par\end{flushleft}{\footnotesize \par}

{\footnotesize{}
\begin{equation}
\Lambda(\sigma,\theta)=\frac{\sigma}{\sqrt{2\pi}\cot i}\exp\left(-\frac{\cot^{2}\theta}{2\sigma^{2}}\right)-\frac{1}{2}\textrm{erfc}\left(\frac{\cot\theta}{\sigma\sqrt{2}}\right)
\end{equation}
}{\footnotesize \par}

{\footnotesize{}The model takes into account two kinds of first-order
reflections rising from the rough surface $dA$: the specular reflection,
a mirror-like reflection where the observed ray is reflected at the
same angle to the surface normal as the incident ray; and the diffuse
reflection, where the incident ray is scattered in all directions
according to the collective properties of the surface, generally given
by a scattering law. Similarly to \citet{2010Icar..208..548G}, in
the present application of \citeauthor{1998ApOpt..37..130V} model,
we assume that the tilt respects the Lommel-Seeliger law. This law
reproduces the outcome of an absorbing surface exponentially attenuating
the incoming light \citep{2005JRASC..99...92F}. }{\footnotesize \par}

{\footnotesize{}Thus, the radiance due to specular reflection in a
rough medium was derived by \citet{1991PhDT........32N}, and is given
by:}{\footnotesize \par}

{\footnotesize{}
\begin{equation}
L_{rs}=C_{s}\frac{P_{ill+vis}(i,e,\varphi,\sigma)}{\cos e\cos^{4}\theta_{a\ spec}}\exp\left(\frac{-\tan^{2}\theta_{a\ spec}}{2\sigma^{2}}\right)
\end{equation}
}{\footnotesize \par}
\noindent \begin{flushleft}
{\footnotesize{}where the $C_{s}$ is a normalizing factor:}
\par\end{flushleft}{\footnotesize \par}

{\footnotesize{}
\begin{equation}
C_{s}=\frac{1}{4\sqrt{\pi}\textrm{U}(-1/2,0,(2\sigma^{2})^{-1})}
\end{equation}
}{\footnotesize \par}
\noindent \begin{flushleft}
{\footnotesize{}and $\theta_{a\ spec}$ is the tilted angle regarding
the specular cone:}
\par\end{flushleft}{\footnotesize \par}

{\footnotesize{}
\[
\theta_{a\ spec}=\arccos\left\{ \left(\cos i+\cos e\right)\right.\left[\left(\cos\varphi\sin e+\sin i\right)^{2}\right.
\]
}{\footnotesize \par}

{\footnotesize{}
\begin{equation}
\left.\left.\sin^{2}\varphi\sin^{2}e+\left(\cos i+\cos e\right)^{2}\right]^{-1/2}\right\} 
\end{equation}
}{\footnotesize \par}

\emph{\footnotesize{}U(a, b, z)}{\footnotesize{} in the normalizing
factor $C_{s}$ is the confluent hypergeometric function. $U(-1/2,\ 0,\ 1/x^{2})$
can be approximated to }{\footnotesize \par}

{\footnotesize{}
\begin{equation}
U(x)=\frac{1}{\sqrt{\pi}}\frac{\sqrt[2x^{2}]{e}}{2x^{2}}\left[K_{0}\left(\frac{1}{2x^{2}}\right)+K_{1}\left(\frac{1}{2x^{2}}\right)\right]
\end{equation}
}{\footnotesize \par}
\noindent \begin{flushleft}
{\footnotesize{}in case }\emph{\footnotesize{}U}{\footnotesize{} is
not numerically available. $K_{n}$ is the modified Bessel function
of second kind.}
\par\end{flushleft}{\footnotesize \par}

{\footnotesize{}The radiance due to the diffusive reflection for every
surface element that is visible and illuminated is obtained by numerically
integrating over all tilted $\theta_{a}$ and $\varphi_{a}$:}{\footnotesize \par}

{\footnotesize{}
\[
L_{rd}=P_{ill+vis}(i,e,\varphi,\sigma)\int_{0}^{\pi/2}\left[\int_{a}^{b}\frac{2\cos\theta_{i}^{'}}{\cos\theta_{r}^{'}+\cos\theta_{i}^{'}}\frac{\textrm{d}\varphi_{a}}{2\pi}\right]\times
\]
}{\footnotesize \par}

{\footnotesize{}
\begin{equation}
\times\frac{\cos\theta_{i}^{'}}{\cos\theta_{a}\cos e}P_{a}(\theta_{a},\sigma)d\theta_{a}
\end{equation}
}{\footnotesize \par}
\noindent \begin{flushleft}
{\footnotesize{}where $\theta_{r}^{'}$ and $\theta_{i}^{'}$ are
the modified incidence and emergence angles by the local tilted surface
and given by:}
\par\end{flushleft}{\footnotesize \par}

{\footnotesize{}
\begin{equation}
\cos\theta_{i}^{'}=\cos\varphi_{a}\sin i\sin\theta_{a}+\cos i\cos\theta_{a}
\end{equation}
}{\footnotesize \par}

{\footnotesize{}
\begin{equation}
\cos\theta_{r}^{'}=\cos(\varphi_{a}-\varphi)\sin e\sin\theta_{a}+\cos e\cos\theta_{a}
\end{equation}
}{\footnotesize \par}

{\footnotesize{}For the integration limits $a$ and $b$ and their
associated conditions, the reader should see again \citet{1998ApOpt..37..130V}
(Eqs. 9 \& 10 therein) or \citet{2010Icar..208..548G} (Table A1 therein).
When $\sigma=0$, we have $L_{rd}\rightarrow\cos i/(\cos i+\cos e)$,
the Lommel-Seeliger law.}
\begin{figure*}[!t]
\includegraphics[scale=0.35]{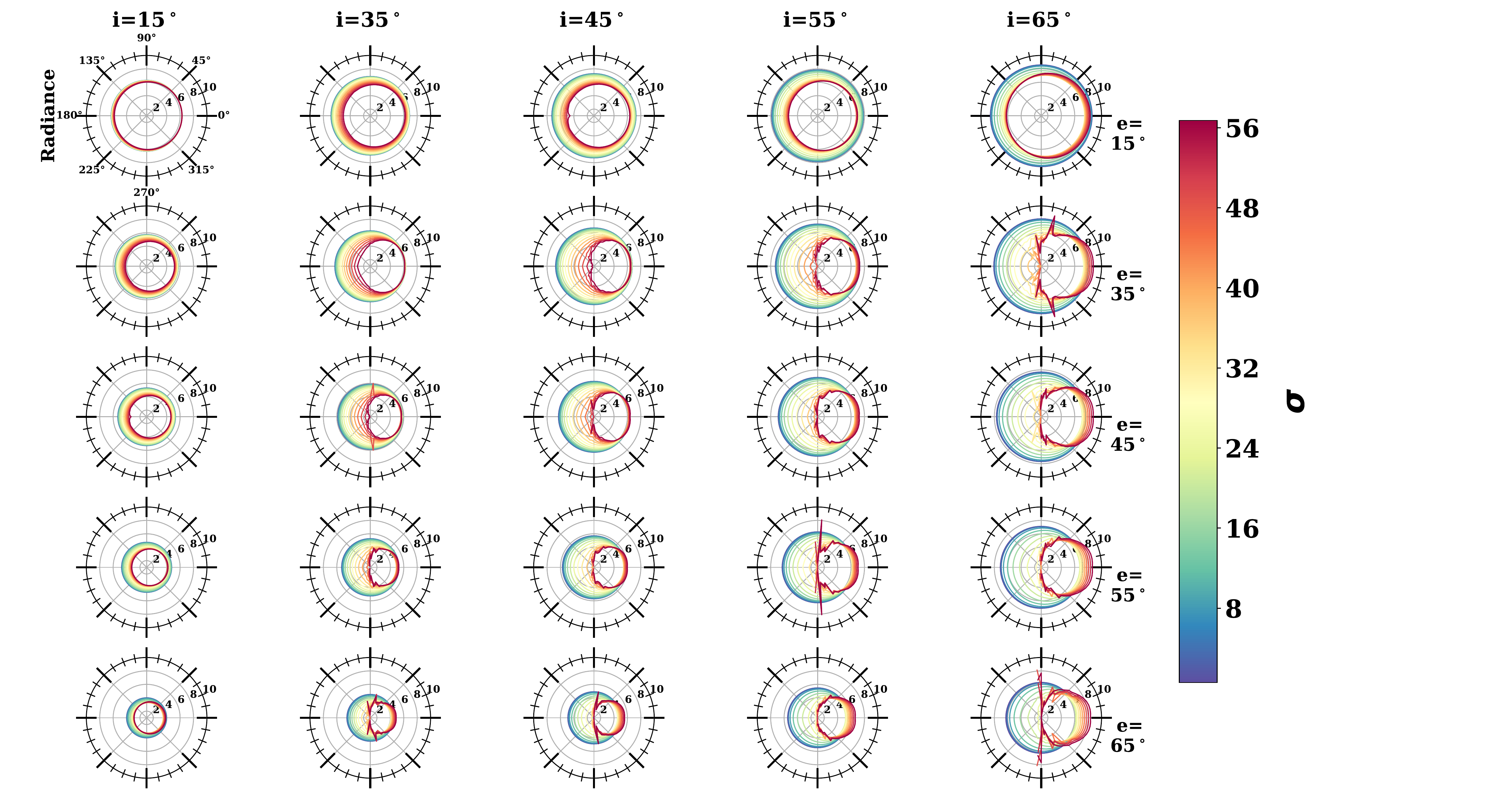}

{\footnotesize{}\caption{\label{fig:Lrd_profiles}{\footnotesize{}$L_{rd}$ radiance in azimuthal
polar profiles. Each color-coded profile is linked to a given roughness
RMS $\sigma$. For every column the $i$ angles are fixed, while the
$e$ angles increases along the rows.}}
}{\footnotesize \par}

\end{figure*}
{\footnotesize{} In Figure \ref{fig:Lrd_profiles}, the polar $L_{rd}$
profiles show how the function becomes increasily dominated by backscattering
as $i,\ e,$ and $\sigma$ gets higher, i.e., roughness increases
the incident radiance that is scattered back over the observer. At
low $\sigma$, the function is nearly symmetric in azimuth. Spikes
are observed at increasing roughness and emergence angles, they are
effects of coupling between the titlt distribution and the bright
limb from the Lommel-Seeliger Law.}{\footnotesize \par}

{\footnotesize{}A later addition to the model is the approximative
diffuse inter-reflection contribution among the tilted surfaces. We
assume that the diffuse component is more important than the specular
one. Derived by\citet{1995IJCV.14...227O} also for a Gaussian distribution
of heights, the inclusion of this term is advised by \citeauthor{1998ApOpt..37..130V}
in their 1998 paper. Using the Lommel-Seeliger law, we have the expression:}{\footnotesize \par}

{\footnotesize{}
\[
L_{rd}^{(2)}(i,e,\varphi,\sigma)=0.17\frac{\cos i}{\pi(\cos i+\cos e)}\frac{\sigma^{2}}{\sigma^{2}+0.13}\times
\]
}{\footnotesize \par}

{\footnotesize{}
\begin{equation}
\times\left[1-\left(\frac{\min\left[i,e\right]}{\pi}\right)^{2}\cos\varphi\right]
\end{equation}
}{\footnotesize \par}

{\footnotesize{}Specular, inter-reflection and diffuse radiance contributions
are put together in the final equation for the RADF $L_{r}$:}{\footnotesize \par}

{\footnotesize{}
\begin{equation}
L_{r}(i,e,\varphi,\sigma)=(1-g)\cdot\rho\cdot p_{sca}(\alpha)\left[L_{rd}+\rho\cdot L_{rd}^{(2)}\right]+g\cdot L_{rs}
\end{equation}
}{\footnotesize \par}
\noindent \begin{flushleft}
{\footnotesize{}$\rho$ is the approximative single-scattering albedo;
$p_{sca}(\alpha)$ is the scattering phase function that accounts
for the wide phase angle-dependence; $g$ is a parameter varying from
0 to 1 balancing the specular and diffuse contribution.}
\par\end{flushleft}{\footnotesize \par}

{\footnotesize{}In this paper, the roughness model was implemented
using Python 2.7.15 and Cython 3.0.0 to speed up calculations \citep{behnel2011cython,Scipy}.
The }\emph{\footnotesize{}U}{\footnotesize{} and $K_{n}$ functions
are available for Python, under the scipy.special package}\footnote{{\footnotesize{}\url{https://docs.scipy.org/doc/scipy/reference/special.html}}}{\footnotesize{}.
The double integrals were evaluated numerically using scipy.integrate.nquad}\footnote{{\footnotesize{}\url{https://docs.scipy.org/doc/scipy/reference/integrate.html}}}{\footnotesize{},
a python wrapping for the Fortran library QUADPACK. To further speed
up the calculations during the data inversion procedure, we interpolate
$L_{rd}$ using scipy.interpolate.GridRegularInterpolator with steps
of $(i,e,\varphi,\sigma)=(3,3,5,2)$ degrees.}{\footnotesize \par}

\subsection{{\small{}Scattering phase function}}

{\footnotesize{}The scattering phase function (SPF) is tightly correlated
to the collective properties of the scatterers that compose the medium
in which we define the rough surface element. Optical constant, size
distribution, and particle shape are the main medium properties when
modeling a particulate surface \citep{1994JQSRT..52...95M,2009JQSRT.110..808M,2018JGRE..123.1203I}.
However, in our present approach to treating the phase function, we
focus only on retrieving the general shape of this function. The shape
can be compared to more rigorous models in subsequent works \citep{2011A&A...531A.150M,2018ApJ...868L..16M,2018JGRE..123.1203I}.
It is appropriate to notice that multiple scattering is also an important
component even for very dark surfaces. \citet{2001OptSp..91..273Z}
has shown through ray-tracing the polarization of dark carbonaceous
surfaces ($\rho\sim3\%$) requires up to 4 orders of scattering. \citet{2004JQSRT..88..267S}
measured the polarization curve of dark volcanic ash ($\rho\sim10\%$)
in jet stream (``single-scattering'') and deposited modes, finding
significant differences between the two curve slopes due to increasing
multiple scattering from packing. }{\footnotesize \par}

{\footnotesize{}Because our data are out of the opposition effect
regime, we do not incorporate any ad hoc function to separately model
the coherent-backscattering \citep{2009ApJ...705L.118M} nor the shadow-hiding
mechanism \citep{2015P&SS..118..250W}. If any contribution of the
shadow-hiding mechanism ``leaks'' into the scattering phase function
at intermediary phase angles, we expect the SPF to bundle all these
effects together. It is therefore why we prefer \textquotedblleft scattering
phase function\textquotedblright , instead of assigning the widely-used
``single-particle scattering phase function'' nomenclature of \citet{Hapke2012}.}{\footnotesize \par}

{\footnotesize{}The scattering phase function of an ensemble of packed
particles has generally a bi-lobal shape, with forward and backward
lobes, i.e., towards or away from the observer. The intensity and
relative strength of the lobes are related to average single particle
properties such as transparency, shape and size. An important parameter
is the asymmetric factor, that quantifies the intensity of light scattered
forward (positive value) or backward (negative value) in the phase
function. Therefore, we apply the widely-used bi-lobal Henyey-Greenstein
(HG3) function to model the wider phase angle dependence of the phase
curve \citep{1965ApJ...142.1563I} and provided morphological parameters
for comparison with other solar system bodies. The function is given
as:}{\footnotesize \par}

{\footnotesize{}
\[
p_{sca}(\alpha,b_{1},b_{2},c)=\frac{1+c}{2}\frac{1-b_{1}^{2}}{(1-2b_{1}\cos\alpha+b_{1}^{2})^{^{3/2}}}
\]
 }{\footnotesize \par}

{\footnotesize{}
\begin{equation}
+\frac{1-c}{2}\frac{1-b_{2}^{2}}{(1+2b_{2}\cos\alpha+b_{2}^{2})^{3/2}}
\end{equation}
}{\footnotesize \par}
\noindent \begin{flushleft}
{\footnotesize{}where $b_{1}$ and $b_{2}$ are respectively the backward
and forward lobe widths and $c$ is the relative strength of both
lobes. HG3 is normalized such as $\int_{4\pi}\frac{d\varOmega}{4\pi}p_{sca}=1$.
The asymmetric factor is $\xi=<\cos\theta>=-\frac{1+c}{2}b_{1}+\frac{1-c}{2}b_{2}$.
The $b_{1}$ and $b_{2}$ vary between 0 and 1, while $c$ can go
from -1 (total forward) to 1 (total backward). }
\par\end{flushleft}{\footnotesize \par}

\section{{\small{}Inverse problem}}
\noindent \begin{flushleft}
{\footnotesize{}Our approach to inverting the semi-numerical roughness
model is different to what we have applied for the Hapke Isotropic
Multi-Scattering Approximative model \citep{2016Icar..267..135H,2016MNRAS.462S.287F,2017MNRAS.469S.550H}.
Firstly, we scale the sample size to obtain only the general RADF
profile from the data: we bin the RADF $r_{F}$ data table containing
$(i,e,\alpha,r_{F})$ for every cropped image of the four candidate
sample sites in ($i=25$, $e=25$, $\alpha=10$) bins. The data are
thereby reduced from >1 million to 336,040 points at the interval
of approximately ($3^{\circ}$,$3^{\circ}$,$0.06^{\circ}$). The
azimuth angle is then calculated for every central point and the corners
of the bin through an equation relating $\alpha$ to $\varphi$}\footnote{{\footnotesize{}$\cos\alpha=\cos i\cos e+\sin i\sin e\cos\varphi$}}{\footnotesize{}
\citep{2011P&SS...59.1326S}. Secondly, we run the MCMC twice to sample
the multi-parametric space in order to reconstruct the posterior probability
distribution of solutions for every free parameter, i.e., $(\rho,\sigma,g,b_{1},b_{2},c)$,
from which the statistics for every solution will be estimated.}
\par\end{flushleft}{\footnotesize \par}

\noindent \begin{flushleft}
{\footnotesize{}The MCMC method is inserted in the Bayesian statistics
framework: any }\emph{\footnotesize{}a priori}{\footnotesize{} knowledge
about the initial probability distribution for the free parameters
is taken into account to infer the final }\emph{\footnotesize{}a posteriori}{\footnotesize{}
probability distributions \citep{1995JGR...10012431M,2015Icar..260...73S}.
MCMC promotes controlled random walks through the multi-dimensional
space; exploring it by maximizing the log-likelihood functions. After
a sufficient number of steps, the chain will correspond to the final
probability distributions, independently of any }\emph{\footnotesize{}a
priori}{\footnotesize{} knowledge. The advantages of MCMC are that
the }\emph{\footnotesize{}a posteriori}{\footnotesize{} distributions
are not necessarily normal-like, and that uncertainties and distribution
skewness can therefore be estimated. }
\par\end{flushleft}{\footnotesize \par}

\noindent \begin{flushleft}
{\footnotesize{}The first MCMC run using all free parameters is sampled
at enough steps to constrain the scattering phase function parameters
$(b_{1},b_{2},c)$. On the second run, we fixed $(b_{1},b_{2},c)$
and let it once more reconstruct the distributions for $(\rho,\sigma,g)$.
In our implementation, we computed the chain jumps using the adaptive
Metropolis-Hasting method \citep{haario2001}. We dispatched a chain
of 5000 steps. In the first run, as no previous information is available
about any parameter, we considered }\emph{\footnotesize{}a priori}{\footnotesize{}
uniform probability distributions in the proper range defined for
each parameter (Section 4). For the }\emph{\footnotesize{}a posteriori
}{\footnotesize{}information, we defined two target log-likelihood
functions:}
\par\end{flushleft}{\footnotesize \par}
\begin{itemize}
\item {\footnotesize{}In the first run, MCMC tries to fully match the $L_{r}$
distribution to the $r_{F}$ distribution. For every step of the chain,
we compute the Kernel Density Estimator \citep{1992WS...1..S} of
$L_{r}(\rho,\sigma,g,b_{1},b_{2},c)$ distribution as it maximizes
the log-likelihood in respect to data $r_{F}$. We expect to better
retrieve the scattering function parameters $(\rho,b_{1},b_{2},c)$
dominating the phase curve.}{\footnotesize \par}
\item {\footnotesize{}In the second run, the distribution of $L_{r}(\sigma,g)/p_{sca}^{'}$
is compared to $r_{F}/p_{sca}^{'}$ , where $p_{sca}^{'}$ is the
scattering phase function in respect to the best solution from $(b_{1},b_{2},c)$
}\emph{\footnotesize{}a posteriori }{\footnotesize{}distributions.
The same procedure as in the first run is applied here. As we remove
the wide phase angle dependence, we expect to better constrain $(\sigma,g)$.}{\footnotesize \par}
\end{itemize}
{\footnotesize{}In the final step, we calculated the autocorrelation
for every parameter, as well as their }\emph{\footnotesize{}a posteriori}{\footnotesize{}
probability distributions and corresponding statistics (i.e., median,
mean, mode, variance, and interquartile ranges). The autocorrelation
informs us whether the parametric space was fully explored. The }\emph{\footnotesize{}a
posteriori}{\footnotesize{} distributions inform us of the probability
that a given solution matches the data. Multi-modality in the }\emph{\footnotesize{}a
posteriori}{\footnotesize{} distribution shows that other solutions
also have a certain probability to describe the data given the applied
model. The final }\emph{\footnotesize{}a posteriori}{\footnotesize{}
distributions were estimated using the Kernel Density Estimator with
the bandwidth given by a Silverman's Rule ($(n\cdot(d+2)/4)^{-1/(d+4)}$,
where $n$ is the number of points and $d$ is the number of dimensions,
\citealp{silverman1986density}).}{\footnotesize \par}

\section{{\small{}Results}}

\subsection{{\small{}MCMC evaluation}}

{\footnotesize{}}
\begin{figure*}[t]
\includegraphics[scale=0.4]{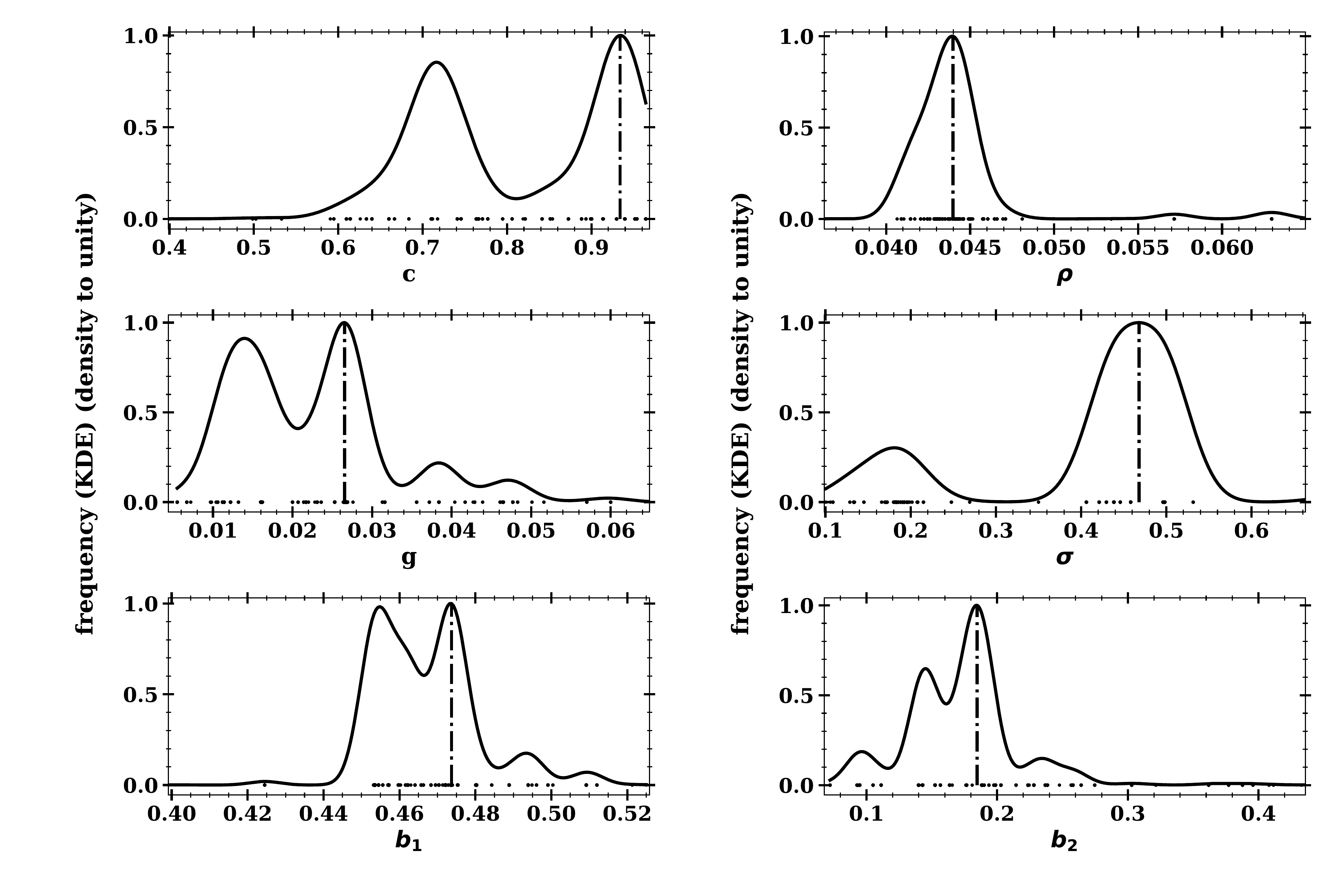}

{\footnotesize{}\caption{\label{fig:KDE_posteriori_dist}{\footnotesize{}Parametric }\emph{\footnotesize{}a
posteriori}{\footnotesize{} distributions from MCMC inversion of the
x-filter RADF data. The MCMC chain values are shown as black dots
sitting where Y-axis is equal to zero. Every distribution is normalized
to density and re-scaled by the mode (dotted straight line). The y-axis
represents the solution probability with respect to the maximum frequency
value. Note: $\sigma$ is in radians.}}
}{\footnotesize \par}

\end{figure*}
{\footnotesize{}Our default analysis was conducted in the x-filter
(847 nm) RADF data. This filter was chosen to facilitate comparison
with a future OVIRS analysis at the same wavelength range. In what
follows, we discuss the parameters with respect to x filter only.
The spectro-photometry and the parameters obtained for the other filters
are presented in Section 6.2.}{\footnotesize \par}

{\footnotesize{}The parametric }\emph{\footnotesize{}a posteriori}{\footnotesize{}
distributions for the x-filter data from MCMC inversion are shown
in Figure \ref{fig:KDE_posteriori_dist}. We can distinguish multi-modal
solutions for many of the parameters, while $c$, $g$ and $b_{1}$
have more pronounced bi-modality. Taking only the first mode and its
associated midspread (IQR, interquartile range), i.e. the difference
between the upper ($Prob(75\%)$) and lower ($Prob(25\%)$) quartiles,
we obtain $\rho=4.4_{-0.2}^{+0.1}\%$, $\sigma=27_{-5}^{\circ+1}$,
$g=2.6_{-0.8}^{+0.1}\%$, $b_{1}=0.470_{-0.004}^{+0.003}$, $b_{2}=0.18_{-0.04}^{+0.01}$,
and $c=0.93_{-0.08}^{+0.07}$ as the most probable solution. A second
mode is found at $\sigma=11_{-6}^{\circ+3}$, $g=1.5_{0.1}^{1.8}\%$,
$b_{1}=0.455_{-0.005}^{+0.003}$, and $c=0.71_{0.6}^{0.8}$. }{\footnotesize \par}

\subsubsection{\textsf{\textsl{\footnotesize{}Roughness patterns in the reflectance}}}

{\footnotesize{}}
\begin{figure*}[!t]
\begin{centering}
\subfloat[\ ]{\begin{centering}
\includegraphics[scale=0.32]{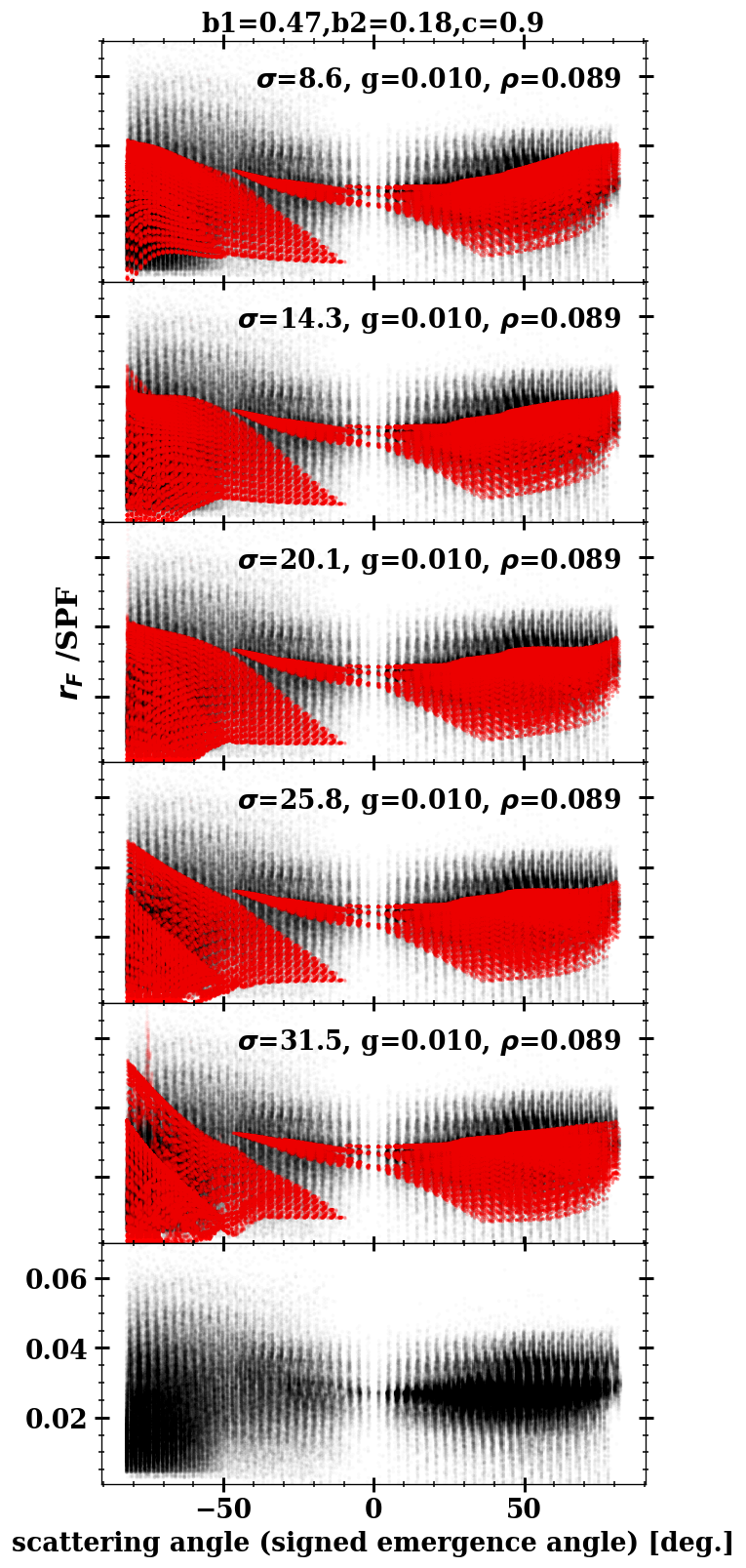}
\par\end{centering}
}\subfloat[\ ]{\begin{centering}
\includegraphics[scale=0.32]{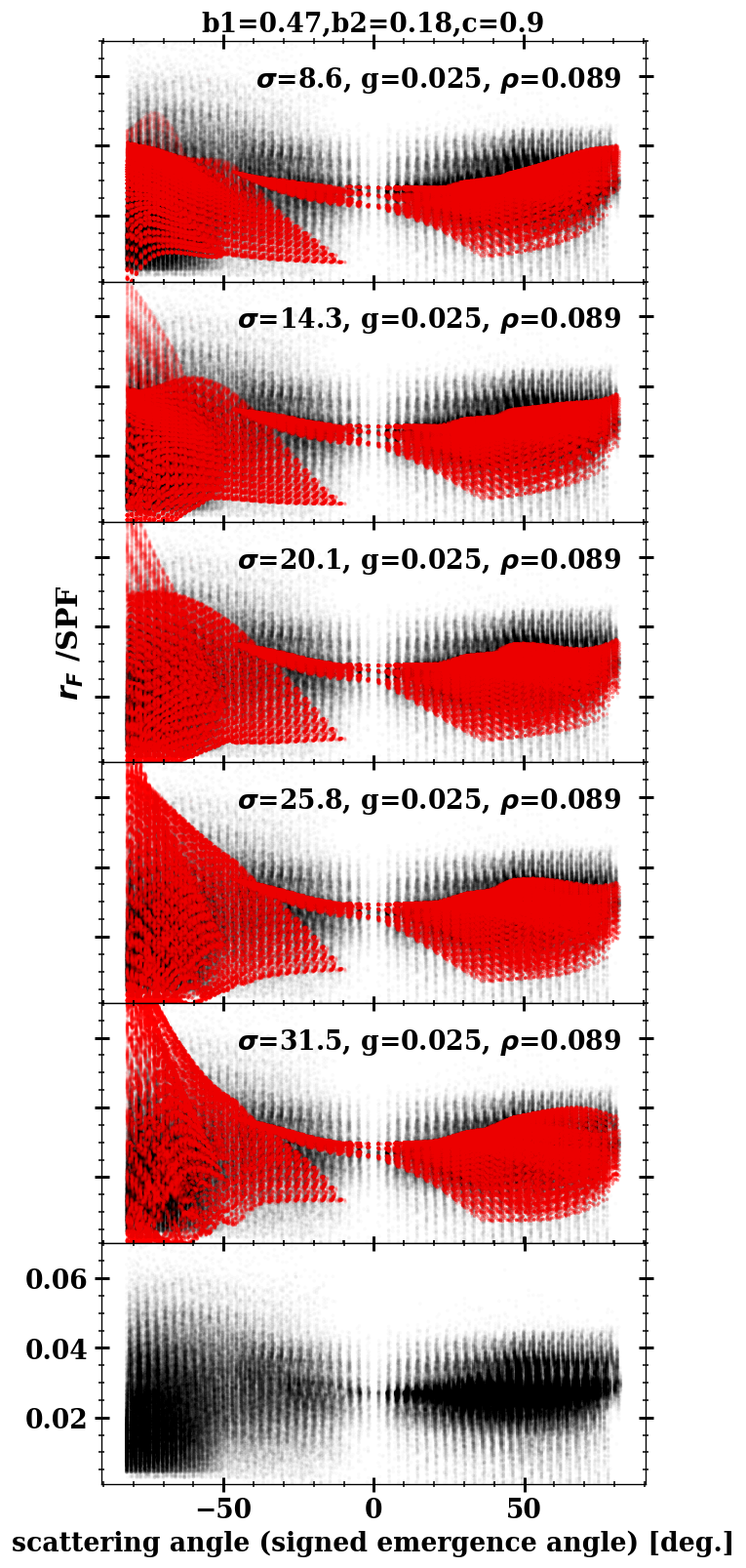}
\par\end{centering}
}\subfloat[\ ]{\begin{centering}
\includegraphics[scale=0.32]{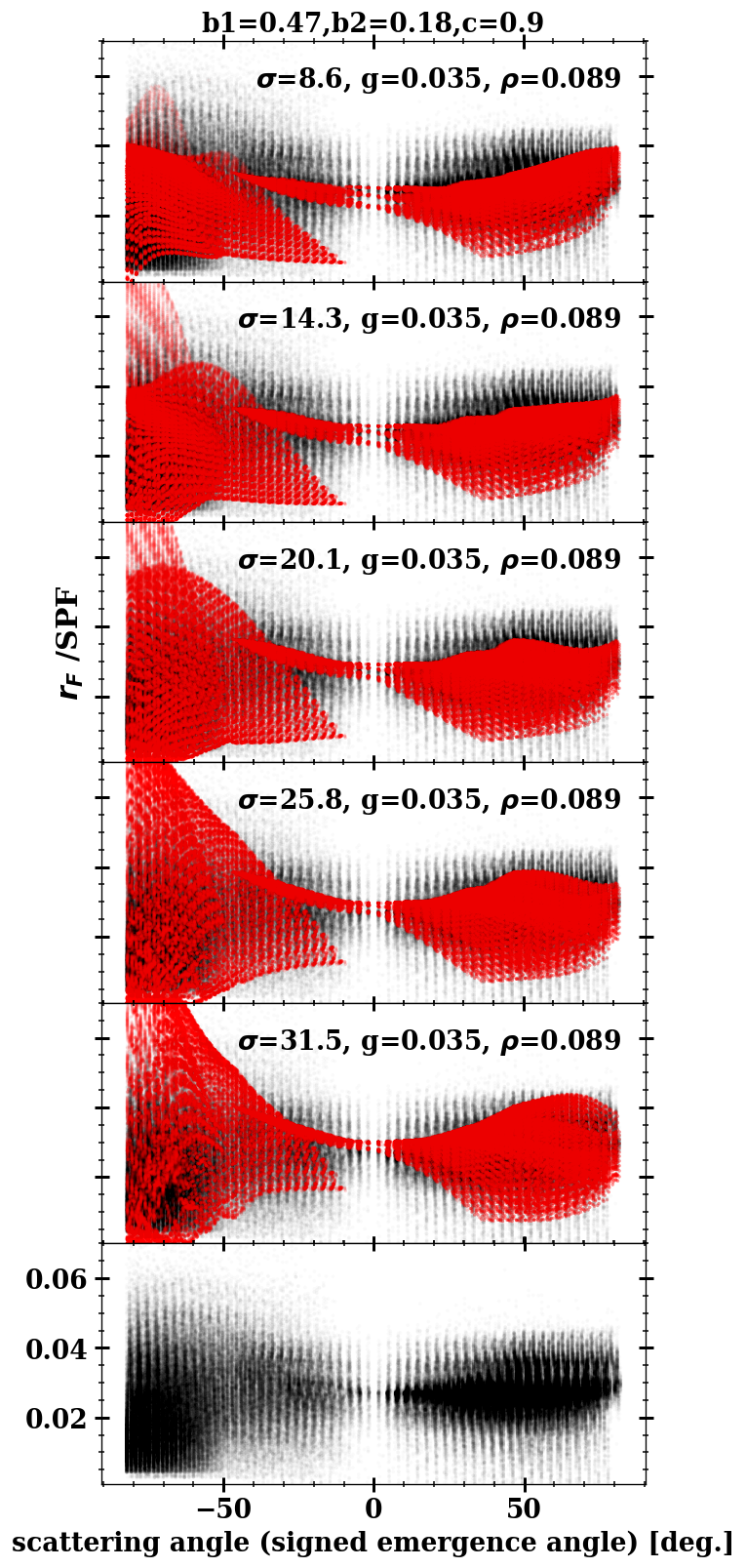}
\par\end{centering}
}
\par\end{centering}
{\footnotesize{}\caption{\label{fig:sca_profile1}{\small{} }{\footnotesize{}Scattering profiles
for the $r_{F}$ and $L_{r}$ distributions in function of the signed
emergence angle. The profiles were calculated for $\sigma=\{8.6^{\circ},14.3^{\circ},20.1^{\circ},25.8^{\circ},31.5^{\circ}\}$
at three different values of specular ratio: (a) $g=1\%$, (b) $g=2.5\%$,
and (c) $g=3.5\%$. $L_{r}$ distributions are in red, while $r_{F}$
distributions are in black. All panels are constrained to the same
RADF and scattering angle intervals. The first-mode solution is situated
between the fourth and fifth subpanels of the second column.}}
 }{\footnotesize \par}
\end{figure*}
{\footnotesize{}To evaluate the capability of the semi-numerical roughness
model to describe the data variance, we devised an alternative fashion
to visualize $r_{F}$ and $L_{r}$ distribution. We first remove the
wider phase angle dependence by dividing the $r_{F}$ by the SPF calculated
from the first-mode solution. Secondly, we split the scattering geometry
space in two hemispheres: for emergence angles with associated $\varphi>90^{\circ}$,
we assign a minus sign, while for those associated to $\varphi<90^{\circ}$,
a positive sign is assigned. We then separate the measurements obtained
at the forward-scattering configuration from those obtained at the
backward-scattering configuration. Excesses and point agglomerations
at either configuration can be better perceived. The results are shown
in Figure \ref{fig:sca_profile1}.}{\footnotesize \par}

{\footnotesize{}The three panels in Figure \ref{fig:sca_profile1}
illustrate the roughness patterns in the RADF distribution as the
roughness slope and specular factor increase. In panel \ref{fig:sca_profile1}a,
where the $g$ is low, increasing roughness leads to flat forward-scattering
and steeper backward-scattering as a function of the emergence angle.
The agglomeration of forward-scatter faint $r_{F}$ points at high
emergence angles ($\approx-50^{\circ}$) is better covered by a low
$\sigma$. They are more Lommel-Seeliger scatters, which is explains
the secondary MCMC modal solution at $g=1.5\%$ and $\sigma=11^{\circ}$.
This secondary solution indicates that the MCMC walker recognizes
that agglomeration of points is described by other parameters rather
than the ``global solution''. In panel \ref{fig:sca_profile1}c,
the RADF is very sensitive to the increase of a few percent in the
specular ratio. Higher $g$ leads to an increase in data variance
and also a RADF increase in the $-40<e<0$ range. In this model, we
can explain most of the high backscatter dispersion by a rough surface
with a non-negligible specular contribution. }{\footnotesize \par}

{\footnotesize{}The most frequent solution is situated between the
fourth and fifth subpanels ($\sigma=20.1^{\circ}$ and $25.8^{\circ}$,
respectively) of column \ref{fig:sca_profile1}b. This solution covers
most of the forward-scattering distribution, as well as the highest
and lowest RADF points in the backscatter configuration. The modeling
covers most of the data variance in the phase angle, with an average
residual $\left|r_{F}-L_{r}\right|/r_{F}<0.007$. }{\footnotesize \par}

\subsubsection{\textsf{\textsl{\footnotesize{}Scattering phase function lobes $b_{1}$
, $b_{2}$ \& $c$}}}

\textbf{\footnotesize{}}
\begin{figure}
\textbf{\includegraphics[scale=0.3]{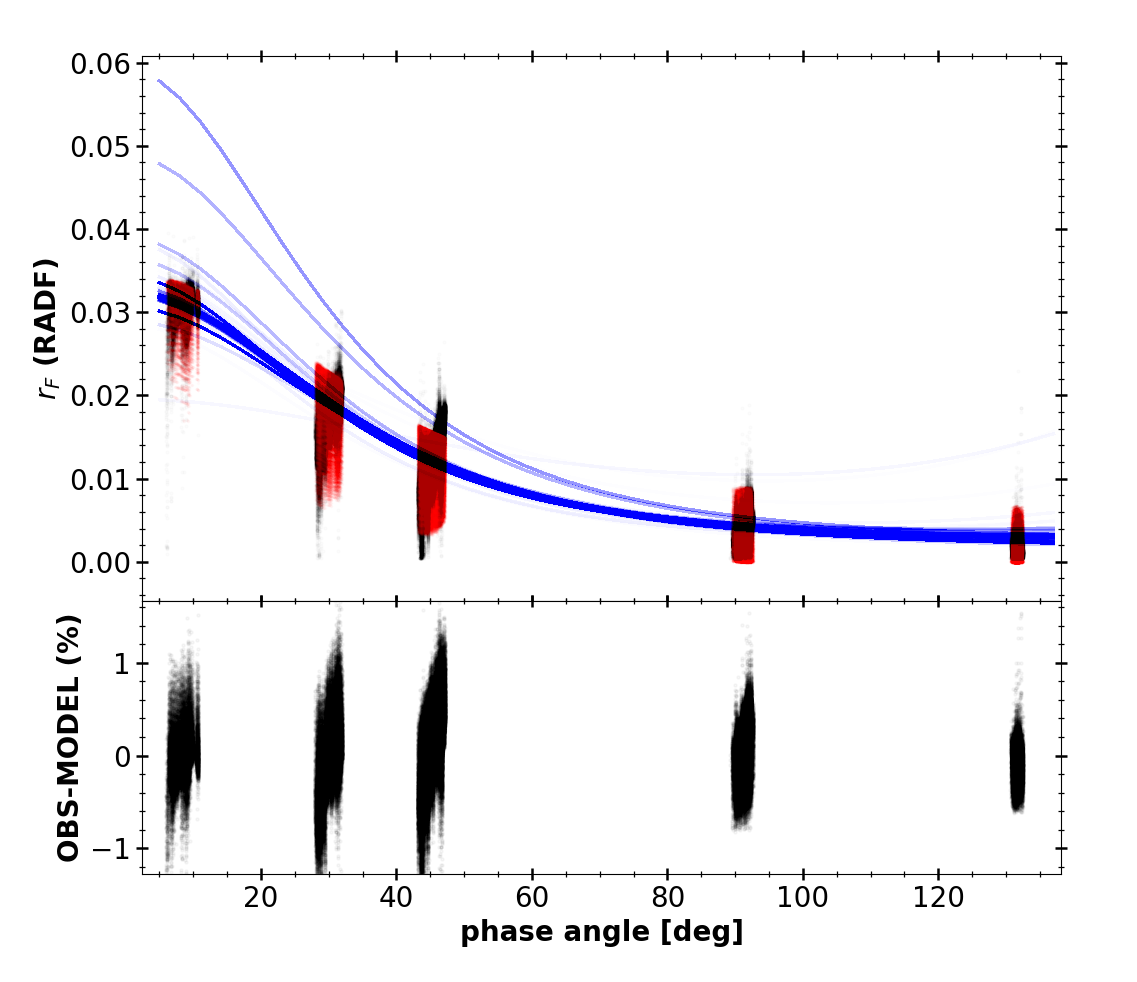}}

\textbf{\footnotesize{}\caption{\label{fig:sca_phasefunc}{\footnotesize{}MapCam x-filter $r_{F}$
distribution in function of the phase angle. In the top panel, the
black points represent all $r_{F}$ data under all DTMs. The superimposed
red points represent the calculated $L_{r}$ distribution from the
first-mode solution. The blue lines are the HG3 SPFs calculated from
all $(b_{1},b_{2},c)$ 5000-step combinations. In the bottom panel,
the black points are given by the difference between $\left|r_{F}-L_{r}\right|/r_{F}$,
in percent.}}
}{\footnotesize \par}
\end{figure}
{\footnotesize{}$b_{1}$ and $c$ parameters indicate that the SPF
is predominantly back-scattering ($\xi=-0.444$, as calculated by
the formula in the section 4.2) in the phase angle range between of}\textbf{\footnotesize{}
$7.5^{\circ}-130^{\circ}$.}{\footnotesize{} $b_{2}$ is smaller,
but hints at a weak forward-scattering lobe at very large phase angle;
however, more data are needed to characterize this scattering feature.
The asymmetric factor may only be negative because we lack of neat
detection of a secondary lobe. Negative asymmetric factors are notoriously
an issue when dealing with phase function of small bodies of the solar
system due to observational constrains. Other dark small bodies however
have hinted into similar lack of broad forward-scattering lobes: for
example, when $r_{F}$ measurements for the phase curve of the nucleus
of the comet 67P/Churyumov-Gerasimenko were extended for up to $\alpha=115^{\circ}$,
no signs of a second lobe was yet detected \citep{2017MNRAS.469S.312G}.}{\footnotesize \par}

{\footnotesize{}We can also verify how well the scattering phase function
describes the data. In Figure \ref{fig:sca_phasefunc} we show all
MCMC step SPFs calculated from MCMC steps overplotted on the $r_{F}$
distribution and the optimal $L_{r}$ distribution. Most of the SPFs
calculated from steps cluster well around the most probable solution,
describing the wide phase angle dependence of the $r_{F}$ distribution.
The phase function does not show obvious signs of a rising second
forward-scattering lobe at $\alpha=130^{\circ}$ , this feature is
only hinted in the }\emph{\footnotesize{}a posteriori}{\footnotesize{}
$b_{2}$ distribution as at least \textasciitilde{}0.2 wide. The phase
function becomes flat in the $90^{\circ}-130^{\circ}$ phase angle
range, where the turning point between both lobes is generally situated.
Broad second lobe has been interpreted as the presence of particles
in the larger-than-wavelength size regime with low internal scatterers
in literature \citep{1995Icar..113..134M,Hapke2012}. On the other
hand, \citet{2015JQSRT.150...42Z} show through discrete dipole approximation
of irregular particle agglomerates that the effects of broadening
forward-scattering lobe increases at $\alpha=130^{\circ}$ if the
size distribution of near-wavelength-size particles becomes also becomes
broader and the particles are less absorbing. This might indicate
therefore the presence of small bright scatterers in the surface of
Bennu in the sub-micrometer range. Nonetheless, only rigorous modeling
may reveal some of the grain size properties \citep{1994JQSRT..52...95M,1997JQSRT..57..767M,2011A&A...531A.150M,2015JQSRT.150...42Z,DLUGACH20112068}.}{\footnotesize \par}

\subsubsection{\textsf{\textsl{\footnotesize{}Approximative single-scattering albedo
$\rho$}}}

{\footnotesize{}Small single-scattering albedo $\rho$ in the visible
range is in-line with other B-type asteroids \citep{2010JGRE..115.6005C}.
While we lack data under $\alpha<7.5^{\circ}$ and we do not include
ad hoc opposition effect terms, our estimated $\rho$ is similar to
the reported geometric albedo by \citet{2019NatAs...3..341D}. The
\citeauthor{1998ApOpt..37..130V} model takes into account the back-scattering
increase as the surface gets rougher at intermediary phase angles
(Figure \ref{fig:Lrd_profiles}), mimetizing one of the shadow-hiding
attributes. Yet, \citet{2020Icarus...GOLISH} identify a non-linear
opposition surge of \textasciitilde{}15\% rising under $\alpha<4^{\circ}$.
This could possibly indicate shadow-hiding or a weak coherent-backscattering
effect, which therefore hints to a slightly different value for single-scattering
albedo \citep{2009ApJ...705L.118M,2015P&SS..118..250W}.}{\footnotesize \par}

\subsubsection{\textsf{\textsl{\footnotesize{}Specular ratio $g$}}}

{\footnotesize{}The specular reflection component is non-zero, impling
a not fully diffusive surface, which is generally assumed when modeling
small-body particulate surfaces. Specular reflection is proportional
to the Fresnel or ``mirror'' reflectivity and predominant in metallic
and monocrystalline materials. A specular component in the scattering
process indicates that materials with such properties are possibly
present on the surface. From image inspection and some previous considerations
of Bennu's composition, we suggest two potential explanations of the
non-zero specular ratio: (i) Some eroding processes may lead to very
flat clean-cut mineral faces on exposed boulder surfaces; or (ii)
very small, bright specular inclusions could be present inside Bennu's
rock matrix.}{\footnotesize \par}

{\footnotesize{}Brightness increases associated with flat rock faces
seem ubiquitous on Bennu's surface, but they may only be an effect
of orientation, as argued in \citet{2020Icarus...GOLISH}. Their reflectances
are greatly reduced after applying a photometric-topographic correction,
which indicates that roughness is the main parameter controlling brightness
(Section 6.3). Small bright inclusions, on the other hand, have been
observed in other dark primitive small bodies, including by the contemporaneous
Hayabusa2 mission in the carbonaceous chondrite-like asteroid (162173)
Ryugu. \citet{2019Sci...365..817J} have counted several in images
taken by MASCOT (Mobile Asteroid Surface Scout), and they appear to
be similar to those found in weakly and mildly aqueous altered carbonaceous
chondrites. They can be up to three times as bright as the average
Ryugu surface yet still are not spatially resolved ($\lesssim0.5$
mm). However, the authors were not able to trace the multi-angular
RADF distribution to be able to confirm the specular behavior. Bright
inclusions are also observed on the ROLIS and CIVA images of Philae/Rosetta,
but they are much less abundant \citep{2017Icar..285..263S}. }{\footnotesize \par}

{\footnotesize{}\citet{2019Icar..333..415P} studied a recently fallen
CM2 meteorite, in which both large and small grain size preparations
of the meteoritic sample (called ``chips'' and ``powder'' therein)
indicate the presence of a specular component in the bi-directional
reflectance distribution when observed at intermediary incidence angles.
The preservation of this component even after changing the grain sizes
shows that the specular reflection is arising from a much smaller
size scale.}{\footnotesize \par}

{\footnotesize{}As the specular elements are below the OCAMS spatial
resolution, we are not able to relate the specular ratio to the size,
albedo, and number, nor can we verify a relation to the bright inclusions.
Images taken during reconnaissance of the sample sites may help us
to further investigate the presence of specular bright inclusions.}{\footnotesize \par}

\subsubsection{\textsf{\textsl{\footnotesize{}Roughness RMS slope $\sigma$}}}

{\footnotesize{}The roughness RMS slope $\sigma$ is the parameter
controlling the major part of the RADF multi-angular spread in the
\citet{1998ApOpt..37..130V} model. The $\sigma$ value of $27_{-5}^{\circ+1}$
is very similar to the v-band average roughness slopes $\bar{\theta}$
of other disk-resolved asteroids derived using Hapke shadowing-roughness
model \citep{1984Icar...59...41H}. The asteroids Gaspra (S-type,
\citealp{1994Icar..107...37H}), Eros (Sw-type, \citealp{2004Icar..172..415L}),
Steins (Xe-type, \citealp{2012Icar..221.1101S}), Ryugu (Cb-type,
Tatsumi et al., in prep.), and the cometary nucleus of 67P/C-G \citep{2017MNRAS.469S.550H},
all have $\bar{\theta}$ situated near $28^{\circ}$. These objects
have different sizes, ages, and compositions, but the same optical
roughness slopes may indicate a similar size scale for their irregularities.
Micro-erosions in the space environment \textemdash{} i.e., processes
such as micro-cratering, particle agglutination, and regolith friction
\textemdash{} possibly quickly converge to surface micro-irregularities
on the order of $25^{\circ}-30^{\circ}$.}{\footnotesize \par}

{\footnotesize{}The optical roughness is smaller than the roughness
obtained through thermal infrared modeling ($43\pm1^{\circ}$, \citealp{2019NatAs...3..341D}).
The ``thermal roughness'' is most sensitive to the smaller end of
the spatial scale, i.e., \textasciitilde{}2 cm. This indicates a break
in surface fractality between the optical, acting on the order of
\textasciitilde{}0.1-1 mm \citep{2003Icar..165..414C}, and thermal
centimeter scales. A surface cannot sustain infinite fractality, and
the break could suggest a regime interface from topographic to particle
size irregularities. This is different from what has been observed
on the Moon. \citet{1999Icar..141..107H} have shown, by analysing
spatially resolved Apollo mission images, that lunar soil is consistently
fractal through a decreasing size scale. Lunar regolith, however,
is dominated by particles of a few tens of microns in size, which
may help sustain the fractality for even smaller size scales, while
Bennu shows weak indication of such structure sizes. If fractal roughness
can be used as an indication of micrometric particles, we may have
another discriminant tool to constrain their presence.}{\footnotesize \par}

\subsection{{\small{}Spectro-photometry of the sample site DTM zones}}

{\footnotesize{}}
\begin{figure*}
\begin{centering}
\includegraphics[scale=0.5]{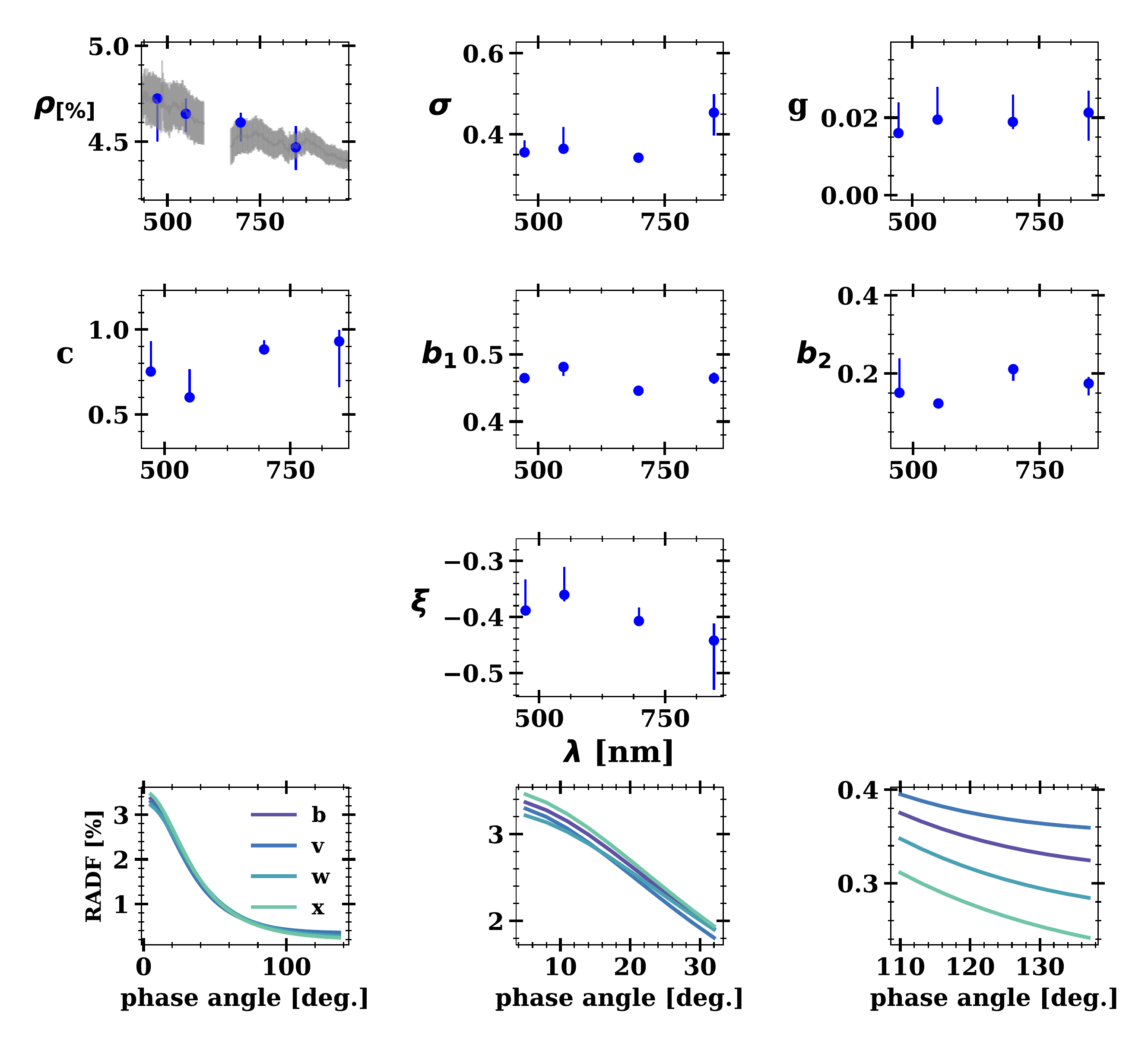}
\par\end{centering}
{\footnotesize{}\caption{{\footnotesize{}\label{fig:spectral_parameters}The spectral behavior
of the surface scattering parameters and the scattering phase function
in the four OCAMS bands. The blue dots represent the mode solution
from their }\emph{\footnotesize{}a posteriori}{\footnotesize{} distributions,
while the errorbars represent the midspread. In the top left panel
(parameter $\rho$) the average Bennu OVIRS EQ3 spectrum segment ($\lambda=$450
\textendash{} 950 nm and $\alpha=7.5^{\circ}$\textendash{} $8.0^{\circ}$)
has been rescaled to match the v-filter albedo (factor of $\times2.17$)
and is superimposed in grey. A segment jump around 660 nm, i.e., where
the spectrum goes from one filter to the next, was removed for clarity.
For the scattering phase functions, the bottom center and bottom right
panels show the zoom-in at small and large phase angles.}}
}{\footnotesize \par}
\end{figure*}
{\footnotesize{}We investigated the spectral behavior of the approximate
single-scattering albedo and other parameters using the same inversion
technique described in Section 5. For all of the OCAMS multi-band
RADF data except those from the x-filter (Section 6.1), we performed
an MCMC evaluation dispatching a chain of 2000 steps. The mode of
the distributions, as well as its midspread for each parameter as
a function of the wavelength, is shown in the Figure \ref{fig:spectral_parameters}.
Heavily skewed error bars indicate the presence of a secondary mode
in the }\emph{\footnotesize{}a posteriori}{\footnotesize{} distribution.
Overall, the surface properties show a weak spectral trend except
in albedo $\rho$ and the asymmetric factor $\xi$. }{\footnotesize \par}

{\footnotesize{}The albedo $\rho$ presents the expected negative
spectral slope ($\varsigma=-0.53\pm0.08\%/\mu m$) related to Bennu's
B spectral asteroid type \citep{2019Natur.568...55L}, and agrees
well with the OVIRS EQ3 global spectral segment in the visible range
taken at $\alpha=7.5^{\circ}$. OVIRS spectra have been radiometrically
calibrated by \citet{2018RemS...10.1486S}. We report an albedo $\rho_{v'}$
of $4.64_{-0.09}^{+0.08}\%$ at 550 nm. The asymmetric factor $\xi$
shows that Bennu becomes more backscattered as wavelength increases,
following same spectral albedo behavior. Trends where $\xi$ is coupled
with $\rho$ have been seen on the surfaces of other dark atmosphereless
bodies, such as Ceres \citep{2019Icar..322..144L} and the nucleus
of 67P/C-G \citep{2015AA...583A..30F}, both observed in the visible
range. In the case of Bennu, the $\xi$ is controlled by the influence
of a second lobe beyond $\alpha>130^{\circ}$, as shown in the bottom
right panel of Figure \ref{fig:spectral_parameters}. The second lobe
weakens and the phase function becomes more back-scattered as wavelength
increases.}{\footnotesize \par}

{\footnotesize{}All of the other surface parameters point to scattering
characteristics already probed through the inversion of the x-filter
data in Section 6.1. The parameters $c$, $b_{1}$ and $b_{2}$ indicate
a backscattering surface with two modal solutions for asymmetric factor
($\xi^{(1)}=-0.360\pm0.030$ and $\xi^{(2)}=-0.444\pm0.020$), and
a possible presence of a weak forward-scattering lobe. The roughness
RMS $\sigma$ ranges between $20^{\circ}$ and $27^{\circ}$ overall
and the specular ratio $g$ seems largely invariant in the visible
range.}{\footnotesize \par}

{\footnotesize{}}
\begin{figure}
\includegraphics[scale=0.6]{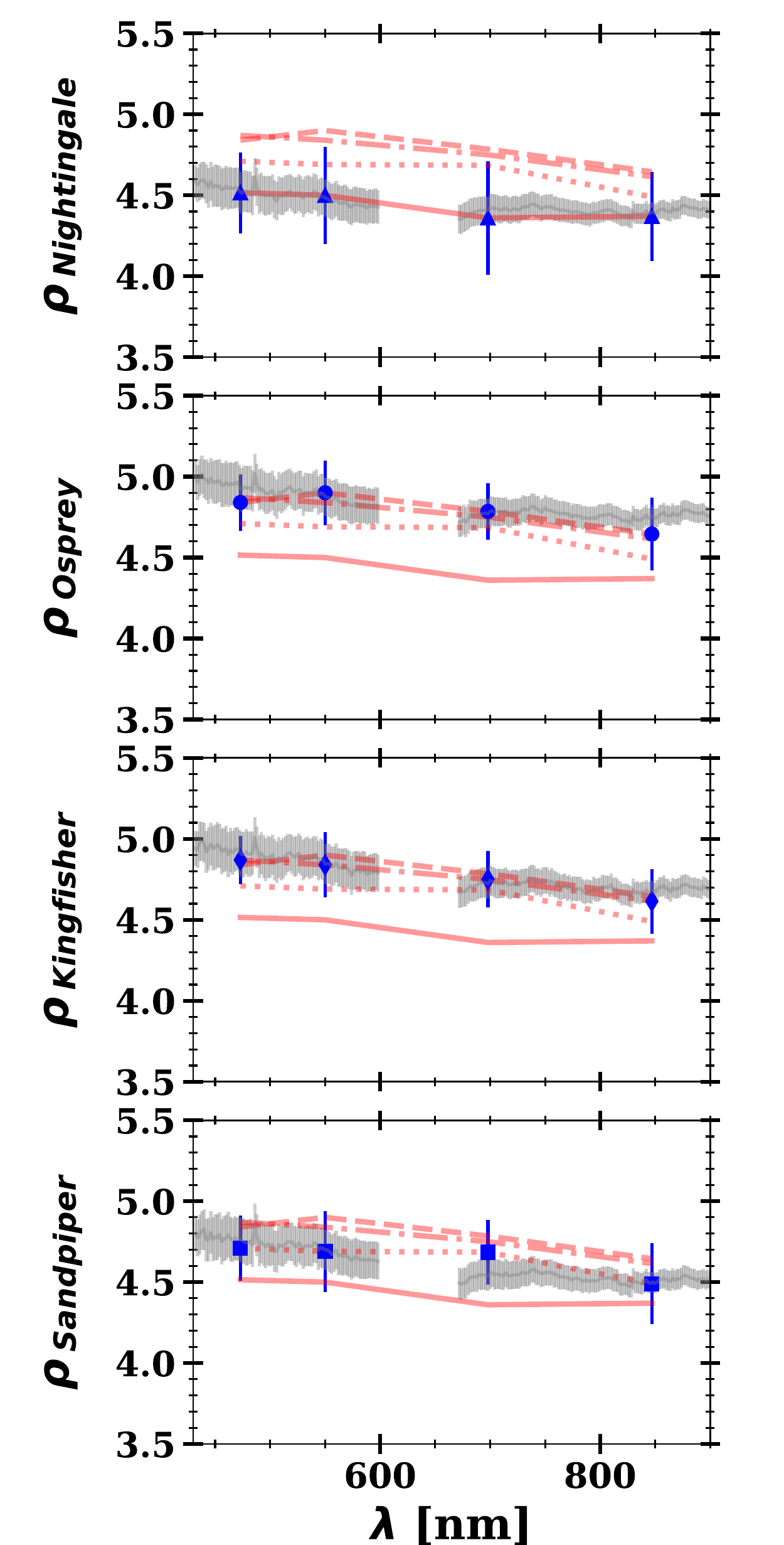}

{\footnotesize{}\caption{\label{fig:sample_site_spectra}{\footnotesize{}The approx. single-scattering
albedo $\rho$ spectro-photometry (blue symbols), in percentage, alongside
the OVIRS EQ3 spectral segment in the visible range (in grey) for
the high-precision DTM zones containing the four sample site candidates.
In each panel, the spectro-photometry of the DTM zones are reproduced
repeatedly in red, for better comparison. The OVIRS were rescaled
to match the $\rho_{v'}$at 550 nm. A segment jump around 660 nm was
removed for clarity. The y-axis and x-axis are fixed to same interval
for clarity. Factors \& Symbols: Nightingale (\textemdash , solid
line) \textemdash{} $\times2.23\%$; Osprey ($--$, dashed line) \textemdash{}
$\times2.32\%$; Kingfisher ($-\cdot$, dashed-dotted line) \textemdash{}
$\times2.23\%$; Sandpiper ($\cdot\cdot\cdot$, dotted line) \textemdash{}
$\times2.15\%$.}}
}{\footnotesize \par}
\end{figure}
{\footnotesize{}We have also investigated the spectro-photometry of
the albedo $\rho$ for each of the four DTM zones containing the sample
site candidates. To obtain their albedo $\rho$, we traced and binned
each of their $r_{F}(\lambda)$ phase curves in the same fashion as
described in the Section 5. We normalized the multi-band phase curves
by dividing them by $L_{r}(\lambda)$ calculated from the optimal
first-mode parameters shown in Figure \ref{fig:spectral_parameters},
leaving only the parameter $\rho$ free. Their approximative single-scattering
albedo spectra are shown in Figure \ref{fig:sample_site_spectra}
alongside their average OVIRS EQ3 spectrum. The site spectra were
averaged for all acquisitions superimposing more than half the nominal
area of the sites. There, as well, we find good agreement between
the $\rho$ spectral trend and the OVIRS EQ3 spectra for the four
sample site DTM zones. Nightingale, which was ultimately chosen as
the primary sample collection site for OSIRIS-REx, is the darkest
and least blue among the four, with $\rho_{v'}^{(N)}=4.5\pm0.06\%$
and spectral slope of $\varsigma^{(N)}=-0.51\pm0.16\%/\mu m$. For
the other candidate sites, we obtain: Osprey, $\rho_{v'}^{(O)}=4.9\pm0.04\%$
and $\varsigma^{(O)}=-0.72\pm0.25\%/\mu m$; Kingfisher, $\rho_{v'}^{(K)}=4.84\pm0.04\%$
and $\varsigma^{(K)}=-0.69\pm0.07\%/\mu m$; and Sandpiper, $\rho_{v'}^{(S)}=4.64\pm0.05\%$
and $\varsigma^{(S)}=-0.70\pm0.27\%/\mu m$.}{\footnotesize \par}

\subsection{{\small{}Photometric correction and the role of roughness}}

{\footnotesize{}}
\begin{figure*}[tp]
\noindent \begin{centering}
(a)\includegraphics[trim=570 0 0 0,clip,scale=0.38]{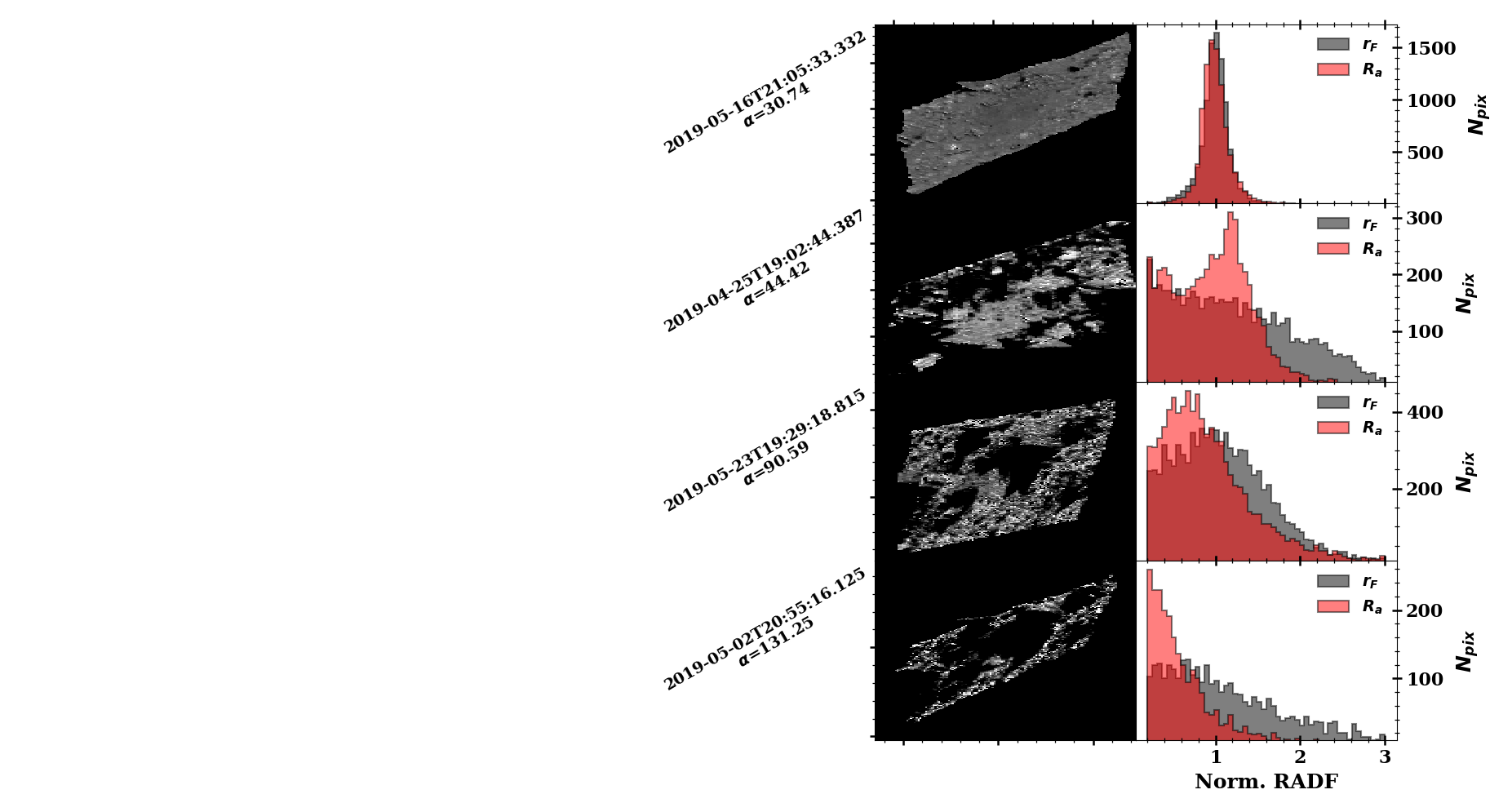}(b)\includegraphics[trim=740 0 0 0,clip,scale=0.38]{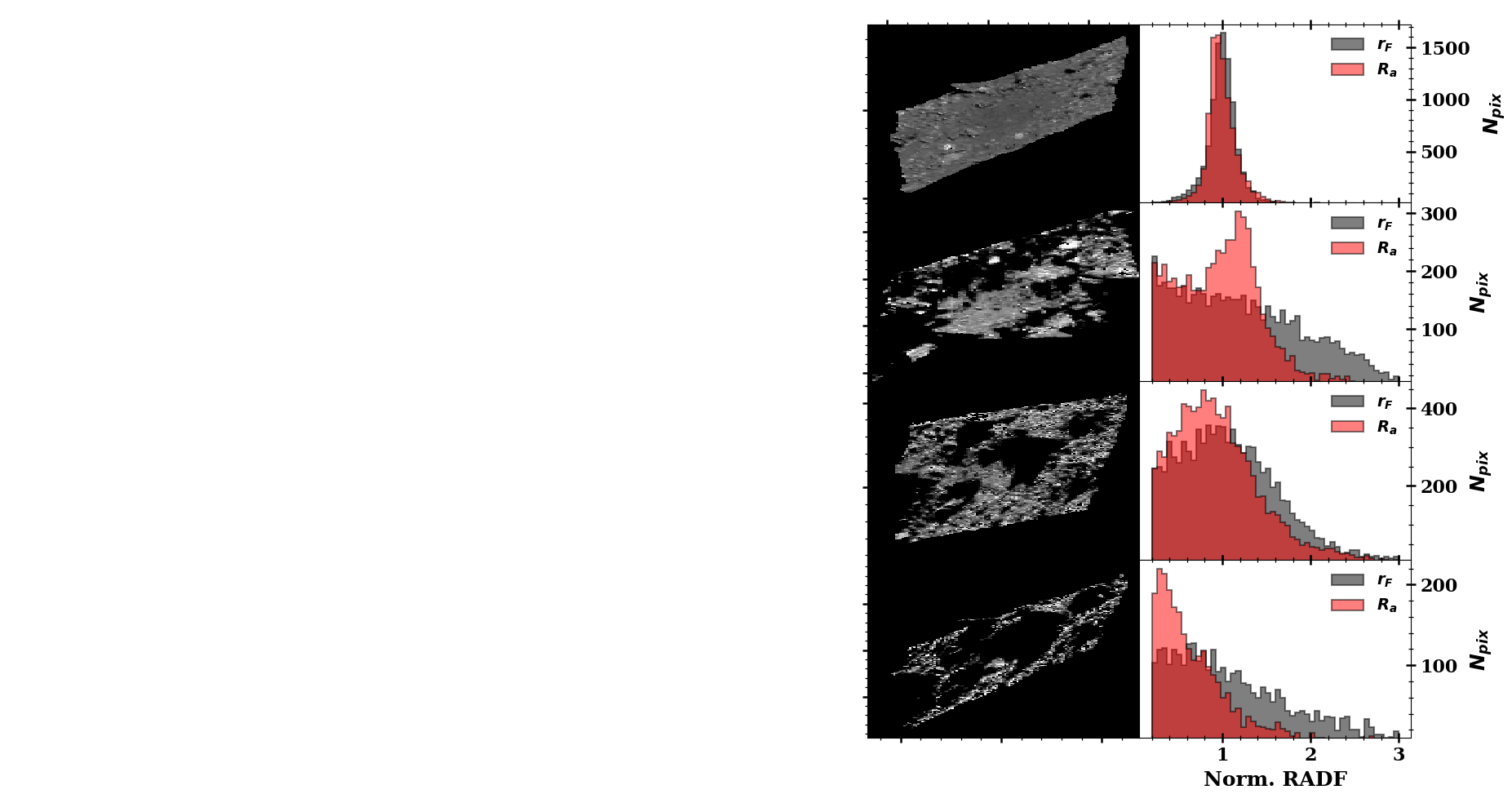}
\par\end{centering}
\noindent \begin{centering}
(c)\includegraphics[trim=570 0 0 0,clip,scale=0.38]{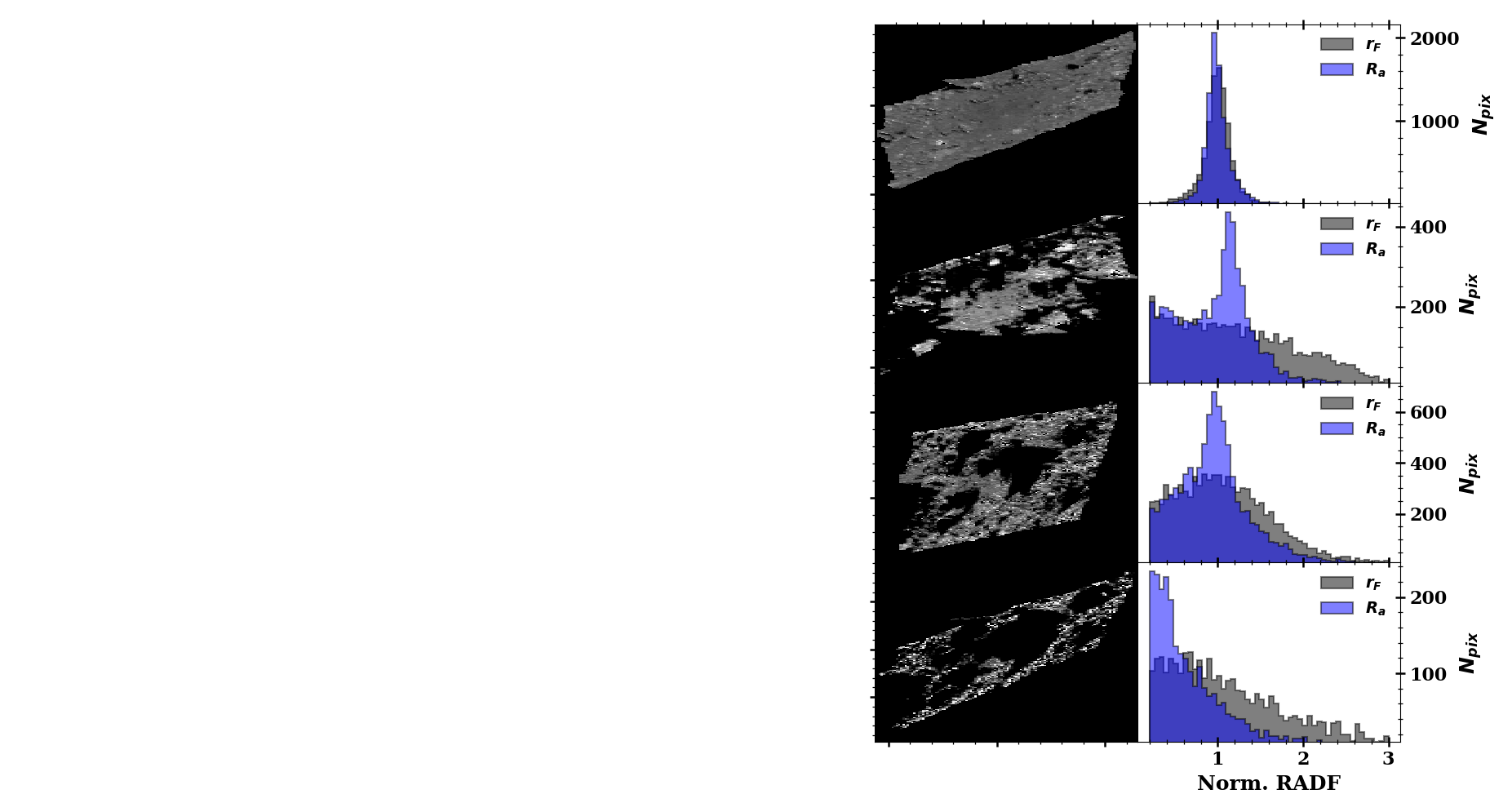}(d)\includegraphics[trim=740 0 0 0,clip,scale=0.38]{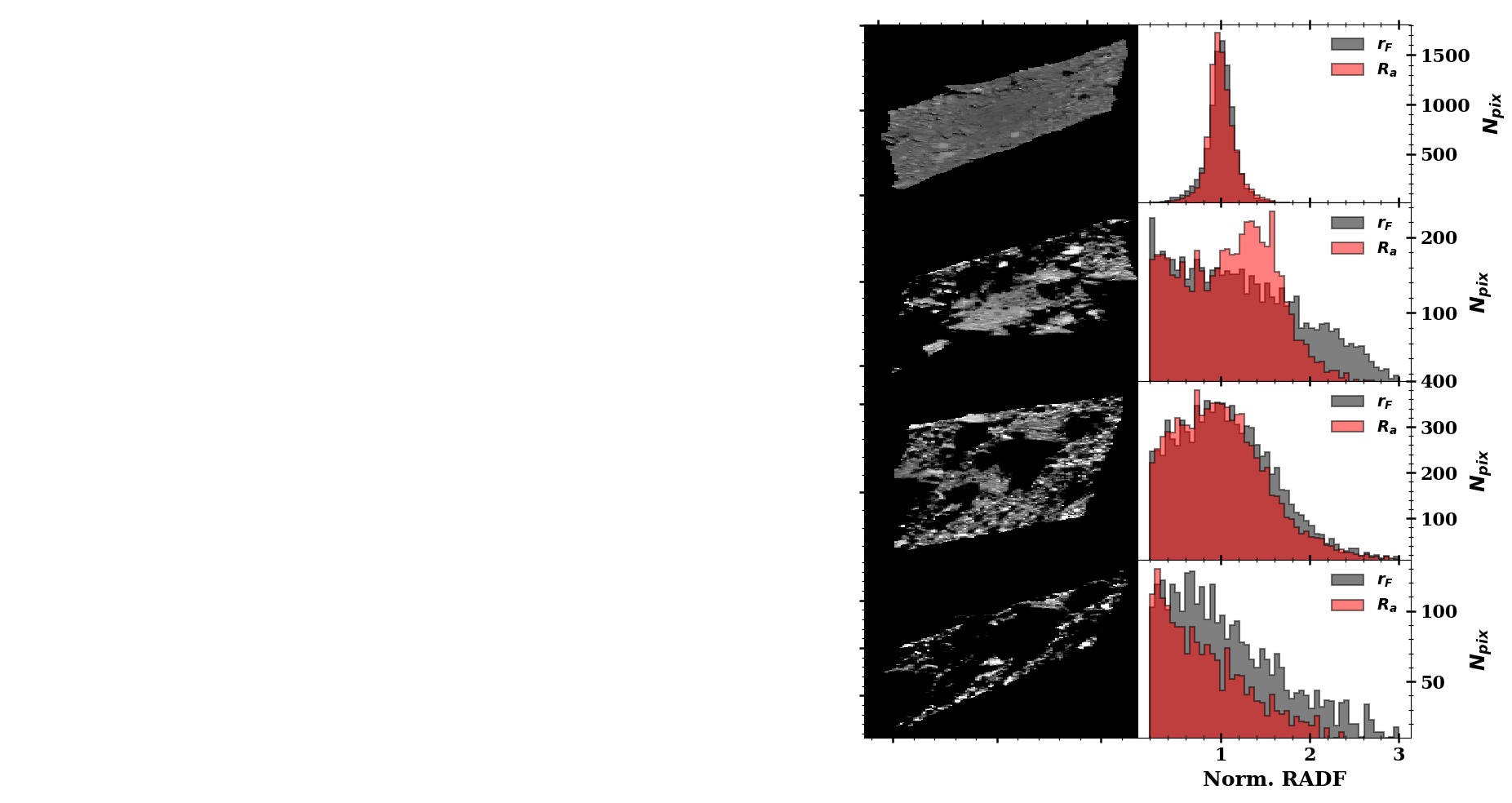}
\par\end{centering}
{\footnotesize{}\caption{{\footnotesize{}\label{fig:photcorr_DL15}Photometric correction of
Nightingale images using mixed solutions. The correction was applied
to four images taken at different phase angles and other observational
conditions. Their timestamps and corresponding phase angles are listed
in the beginning of each row. Each panel corresponds to a different
roughness solution: (a) $\sigma=27^{\circ}$; (b) $\sigma=10^{\circ}$;
(c) a mixture of both solutions; and (d) Lommel-Seeliger correction,
for reference. The first columns of each panel show the corrected
images. Image contrasts and brightness are fixed to same maximum and
minimum levels in all three panels. In the third column of each panel,
the normalized $r_{F}$ histograms of pixels in grey, and the $R_{a}$
histograms representing the photometrically corrected RADF ratio are
shown in red.}}
}{\footnotesize \par}
\end{figure*}
{\footnotesize{}We checked the capacity of the roughness model to
photometrically correct spatially resolved images of a small body
surface. For our tests, we chose four images of the Nightingale site
taken at intermediary to high phase angles. We verified three kinds
of solutions of roughness RMS slope: (a) $\sigma=27^{\circ}$, the
first-mode solution; (b) $\sigma=11^{\circ}$, second-mode solution;
and (c) a mixture of both solutions. All the other parameters were
fixed at the first-mode solution of Section 6.1. }{\footnotesize \par}

{\footnotesize{}We decided to also undertake tests with a lower $\sigma$
motivated by the second mode in the }\emph{\footnotesize{}a posteriori}{\footnotesize{}
distribution (Figure \ref{fig:KDE_posteriori_dist}). This trend is
also evident in Figure \ref{fig:sca_profile1}, where part of the
agglomeration of forward-scatter faint $r_{F}$ points at high emergence
angles are better covered by a low $\sigma$. In the case of mixing
the two solutions, solution (a) or solution (b) was assigned to a
given facet depending on the smallest $\chi^{2}=\sum\left(r_{F}-L_{r}^{(a,b)}\right)^{2}$
residual between the $L_{r}$ from one of two solutions and pixel
$r_{F}$ assigned to the same facet. The facet RADF is divided by
the $L_{r}$ calculated from the appropriate observational condition
using the optimal first-mode solution. In the final step, the shadowed
facets, those ray-traced from the shape model, are removed, and the
corrected RADF ratio $R_{a}$ for every pixel is estimated as described
in Appendix A. }{\footnotesize \par}

{\footnotesize{}}
\begin{figure*}[!t]
\noindent \begin{centering}
\includegraphics[scale=0.5]{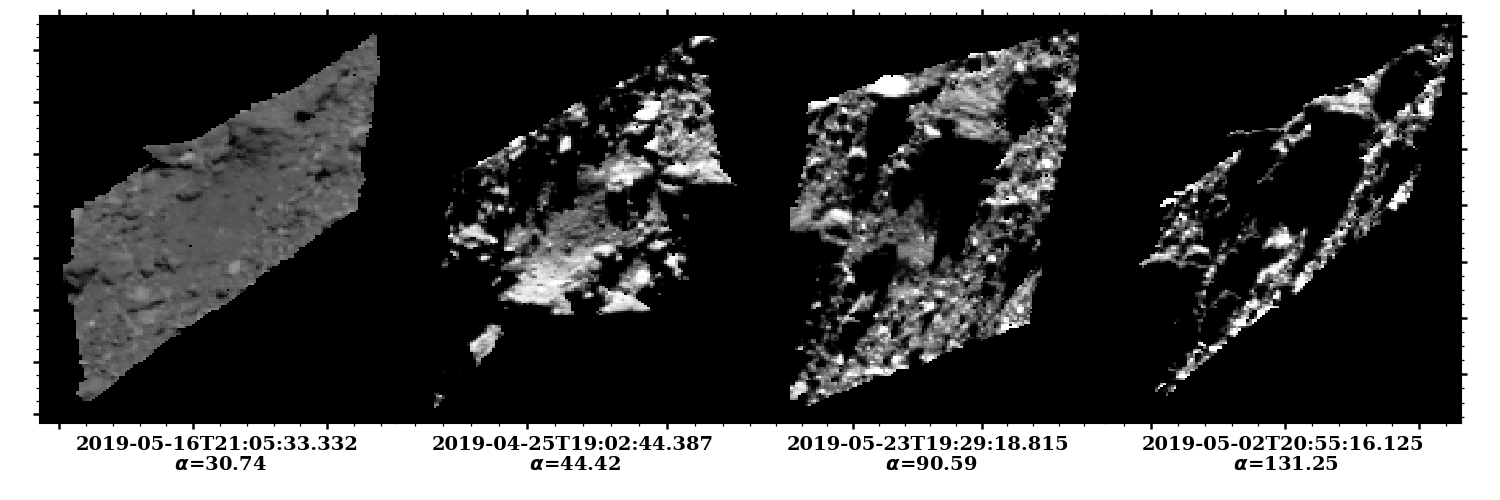}
\par\end{centering}
{\footnotesize{}\caption{{\footnotesize{}\label{fig:Originals}Four EQ image segments of Nightingale
in different phase angles. The contrast and brightness are all scaled
to the same levels as in the Figure \ref{fig:photcorr_DL15}. }}
}{\footnotesize \par}

\end{figure*}
{\footnotesize{}We show in Figure \ref{fig:photcorr_DL15} the results
of those verifications in four intermediary to large phase angles.
For cross-checking, the original image segments of Nightingale, in
the same four different phase angles, are shown in the Figure \ref{fig:Originals}.
A qualitatively good photometric correction is reached when the central
tendency of the corrected RADF $R_{a}$ distribution is unity. This
means that only intrinsic reflectance variation remains in the data.
In Figure \ref{fig:photcorr_DL15}ab, both low and intermediary roughness
slopes yield very similar photometric corrections for intermediary
phase angles ($\alpha\simeq30^{\circ}$ and $\alpha\simeq45^{\circ}$).
In this range, surface roughness is not the main optical factor controlling
reflectance variance, and a Lommel-Seeliger correction is enough to
yield sufficient results. For $\alpha\simeq90^{\circ}$ and $\alpha\simeq130^{\circ}$,
however, the fixed-roughness solutions are insufficient to remove
the photometric-topographic brightness trend, or they overcorrect
it, as in the case of solution (a). This dichotomy between intermediary
and high phase angles is very revealing when we look to the mixed
solutions of Figure \ref{fig:photcorr_DL15}c. By mixing intermediary
and low roughness slopes we obtain a visible improvement in the correction
from $30^{\circ}$ to $90^{\circ}$ phase angle. The apparent bright
topographic structures have their RADF reduced by a factor of up to
3 times at $\alpha\simeq30^{\circ}$ and $\alpha\simeq90^{\circ}$,
indicating that these ``speckles'' are not responsible for the specular
component. For all tested images, the ``replacement ratio'', i.e.,
the ratio of facets with solution (a) to total number of facets, is
about $50\pm5\%$. This ``replacement'' shows no preferential facet
at certain incidence, azimuth, or emergence angles for all Nightingale
images. It means that, at the sub-pixel scale, Bennu\textquoteright s
surface shares two main diffusive components, and only when both components
are taken into account is the photometric correction improved.}{\footnotesize \par}

{\footnotesize{}}
\begin{figure}
\begin{centering}
(a)\includegraphics[trim=425 0 0 0,clip,scale=0.35]{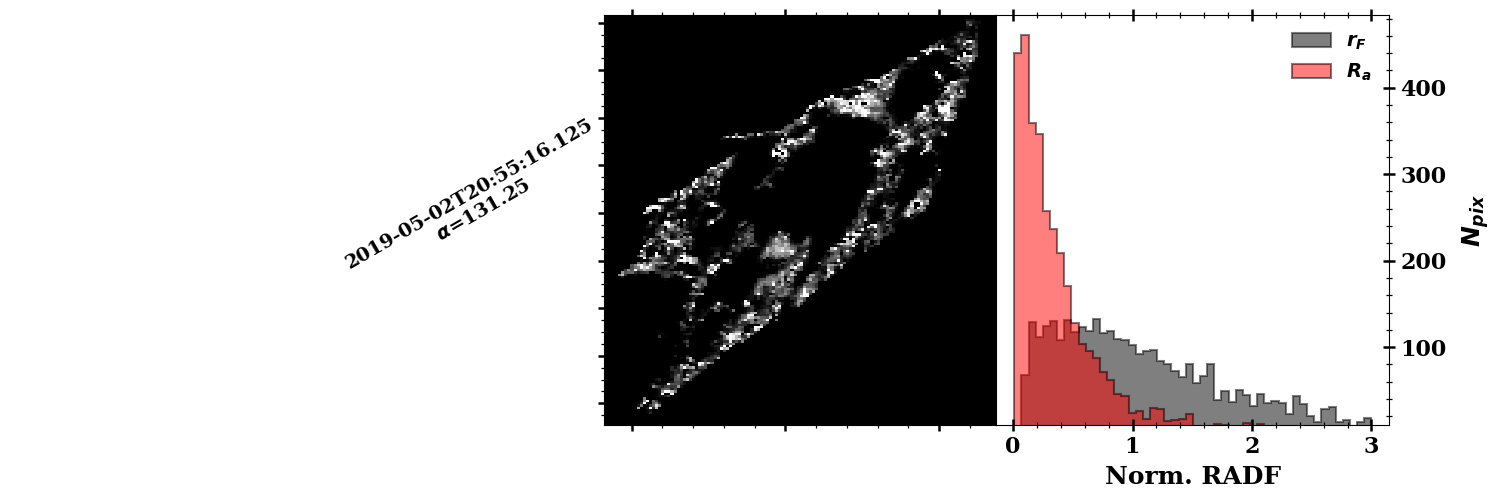}
\par\end{centering}
\begin{centering}
(b)\includegraphics[trim=425 0 0 0,clip,scale=0.35]{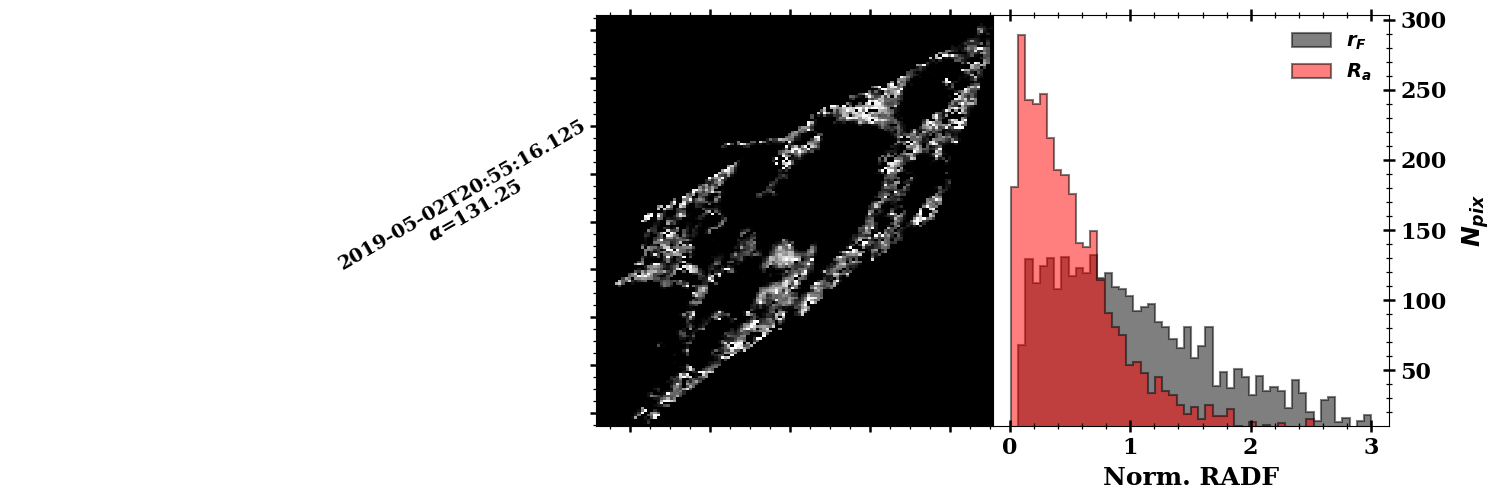}
\par\end{centering}
\begin{centering}
(c)\includegraphics[trim=425 0 0 0,clip,scale=0.35]{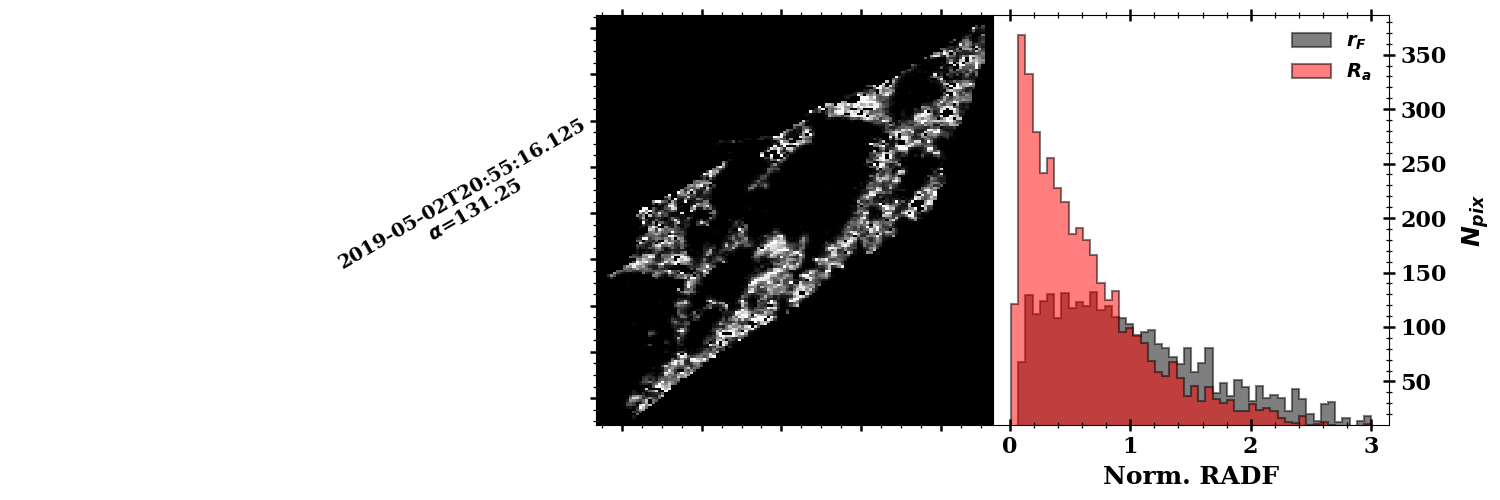}
\par\end{centering}
{\footnotesize{}\caption{\label{fig:photcorr_pha120}{\footnotesize{}Photometric correction
of EQ2 UT 2019-05-02T20:55:16.125 image of Nightingale taken at $\alpha\simeq130^{\circ}$.
(a) Corrected image using $\sigma=55^{\circ}$; (b) same as (a) using
$\sigma=0^{\circ}$; (c) same as (b) without macro-shadow sub-pixel
removal. All images are constrained to same contrast and brightness
levels.}}
}{\footnotesize \par}

\end{figure}
{\footnotesize{}For images taken at $\alpha\simeq130^{\circ}$, the
photometric correction at the sub-pixel level becomes more difficult
to tackle. The images have an average $r_{F}$ of 0.093\% with standard
deviation in a comparable value of 0.1\%, it shows that the reflectance
is more sensitive to topographic features, and therefore, to the shape
model limitations. The mixed solution (Figure \ref{fig:photcorr_DL15}c)
still grants a RADF that is about two to three times as bright as
the data, which indicates that there is another component reducing
even further the RADF at high phase angles. To tackle this, we corrected
the images using $\sigma=55^{\circ}$ and $\sigma=0^{\circ}$, the
upper and lower validity limit of the model, to check how the $L_{r}$
distribution behaves. We also added a third verification, where macro-shadows
are not discounted from the pixel RADF calculation. The effects of
macro-shadows may indicate whether topographic features are playing
a role in the RADF distribution. We present the three verifications
in Figure \ref{fig:photcorr_pha120}. We can first observe that neither
high or very low roughness slope produce enough faint $L_{r}$ values
to correct the data. Figure \ref{fig:photcorr_pha120}c, when compared
to \ref{fig:photcorr_pha120}b, shows that a considerable number of
pixels become fainter if the macro-shadows are not discounted, which
leads the $R_{a}$ distribution to get more skewed to faint levels.
This indicates that, for a high phase angle, a shape model that better
accounts for sub-pixel meso-scale topography is as important as the
sub-millimeter roughness. }{\footnotesize \par}

{\footnotesize{}As the slope distribution of boulder and other topographic
features can be highly non-Gaussian \citep{2017Icar..290...63L},
we propose that mathematically extending the \citet{1998ApOpt..37..130V}
model formulation to non-Gaussian slope distributions might in part
account for the faintness at higher phase angles \citep{1980ITAP...28..788B,1983RaSc...18..566B}.
Another suggestion is to account for fractality \citep{2003JOSAA..20.2081S}
in the diffusive reflection. In this case, the faintness would come
from setting the appropriate fundamental scattering law on the fractal
elements. This could be done using the Akimov disk law \citep{1976SvA...19...385388,1979SvA....23..231A}
as fundamental scattering law or, more extensively, directly computing
the infinite series from \citet{2018Icar..302..213S}, a model that
proposes to describe rough surfaces through multi-scale Gaussian ondulations.
The latter solution may implicate in further computing time and also
departing from the simplicity proposed by the application of \citeauthor{1998ApOpt..37..130V}
framework.}{\footnotesize \par}

\section{{\small{}Discussion and Conclusion}}

{\footnotesize{}We have reintroduced the \citet{1998ApOpt..37..130V}
semi-numerical model to treat first-order light scattering arising
from rough, optically thick surfaces, now coupled with DTM shadow
ray-tracing to account for meso-scale ``rocky'' topography. Our
scientific goal is to provide a parametric description of Bennu's
surface to support laboratory preparations of surface analogs and
the spectral and imaging interpretation of OSIRIS-REx data. We obtained
the scattering parameters and RMS roughness slope of the dark asteroid
Bennu by solving the inversion problem using the MCMC technique applied
to MapCam four-band RADF data for OSIRIS-REx\textquoteright s top
four candidate sample sites together. We also made use of the high-resolution
OLA DTMs produced for these areas of Bennu\textquoteright s surface.}{\footnotesize \par}

{\footnotesize{}The MCMC technique yields}\emph{\footnotesize{} a
posteriori}{\footnotesize{} distributions for each parameter, revealing
interesting aspects of Bennu's surface: while the RMS roughness slope
of $27_{-5}^{\circ+1}$ is in line with what has been obtained for
other asteroids using the Hapke shadowing function, we are puzzled
by the indication of a non-zero specular reflection ratio from the
surface ($2.6_{-0.8}^{+0.1}\%$). The specular reflection hints at
inclusions, possibly of monocrystalline origin, contributing to the
surface reflectance in a way that is generally not taken into account
by fully diffusive approximative radiative transfer models \citep{Hapke2012}.
It may be a direct expression of a compositional sub-centimetric component
on the surface. A plausible candidate for the specular reflection
contribution is carbonate crystal inclusions. Carbonates have been
detected by OVIRS in several areas of Bennu, including in the surrounding
Nightingale (Kaplan et al., \citeyear{2020Scie.....Kaplan}, in press).
Calcites and dolomites appear to be the most abundant component. In
some zones imaged by PolyCam, the carbonates inclusions appear to
be present as bright spots and bright veins of tens of centimeter
size on surface of boulders, whereas the vast majority does not display
any obvious sign of their presence, possibly due to the imaging spatial
resolution. This likely indicates that most of the carbonate is sub-centimetric.
The presence of carbonates, as well as their possible crystalline
phase, provides constraints on the thermal and hydration history of
Bennu and the composition of its parent body (Kaplan et al., \citeyear{2020Scie.....Kaplan},
in press).}{\footnotesize \par}

{\footnotesize{}As for the diffuse rough component, the meso-scale
``rocky'' topography contribution is expected to be pre-modeled
through DTM ray-tracing, leaving the micro-scale roughness \citep{2005Icar..173....3S}
to be described by the \citet{1998ApOpt..37..130V} model. However,
the analysis of the photometric correction of OCAMS images taken at
varied phase angles indicates a more complex scenario. Up to $\alpha\simeq90^{\circ}$,
the photometric correction is greatly improved by mixing two different
solutions for roughness (one with low RMS $\sigma$ and another with
global RMS $\sigma$), a bi-modality already perceived from the MCMC
}\emph{\footnotesize{}a posteriori}{\footnotesize{} distributions.
This bi-modality may indicate the presence of widespread low-roughness
rock faces with quasi-Lommel-Seeliger scattering immersed into other
irregularities account for the broader global distribution of larger
roughness. This kind of landscape is apparently revealed by higher
spatially resolved images taken of the candidate sample sites \citep{2020Icarus...GOLISH}.
We have shown that most of Bennu's brightness variation can be explained
by tuning the roughness slope distribution.}{\footnotesize \par}

{\footnotesize{}Neither the mixing nor pushing the model to its limits
are not enough to yield a satisfactory photometric correction for
images obtained at $\alpha\simeq130^{\circ}$. This points to two
main possible effects: (i) the Lommel-Seeliger scattering law does
not reflect the fundamental diffusive scattering behavior from Bennu's
surface as we approach higher phase angles, which is known to be a
poor law for planetary surfaces \citep{2011P&SS...59.1326S,2018Icar..302..213S}.
(ii) Bennu's tilt distribution is not Gaussian-like at a spatial scale
smaller than the DTM facet size; an over-abundance of small or high
slopes may account for part of the needed faintness. As for the former,
the \citeauthor{1998ApOpt..37..130V} model can be mathematically
adapted to accomodate any scattering law, which is a relevant feature
for future applications, and may reveal which is the actual proper
fundamental scattering law to be used when considering the smallest
unitary tilts in planetary rough surface distributions. As for the
latter, the shadowing can be replicated when further high-resolution
DTMs are available for the candidate sample sites and all of Bennu's
surface. Nonetheless, in future applications of the semi-numerical
model, it may be worthwhile to expand $P_{a}(\theta_{a})$ to non-Gaussian
slope distributions, which may lead to a solution that tackles the
probabilistic terms $P_{ill+vis}$ and $L_{rd}^{(2)}$ through their
full integration \citep{1980ITAP...28..788B,1983RaSc...18..566B,1995IJCV.14...227O,2002ITAP...50..312B}.}{\footnotesize \par}

{\footnotesize{}We report a backscatter scattering phase function
for the phase angle range between $7.5^{\circ}$ and $130^{\circ}$,
without any expressive spectral trend in the visible range. The MCMC
inversion hints at a possible second forward-scatter lobe of at least
\textasciitilde{}0.2 width. This leads to two possible solutions for
the asymmetric factor ($\xi^{(1)}=-0.360\pm0.030$ and $\xi^{(2)}=-0.444\pm0.020$).
We also report a dark global approximate single-scattering albedo
at 550 nm from the collective analysis of all candidate sample sites
of $4.64_{-0.09}^{+0.08}\%$ . The single-scattering albedo from the
MapCam four-band colors has a similar spectral trend to the global
average OVIRS EQ3 spectrum; the four sites together provide a general
description of Bennu's colors. We also find very good agreement in
the spectral slopes between the single-scattering spectro-photometry
and the OVIRS EQ3 spectra of each site candidate separately. }{\footnotesize \par}

{\footnotesize{}On 13 December 2019, Nightingale was announced as
the primary sample site. Although we have not yet conducted a dedicated
photometric analysis of this site, we can provide some predictions
on the surface material structure based on the average photometric
parameters and Nightingale's albedo. Our evidence supports a lack
of widespread sub-micrometric dust, given that the RADF distribution
is sufficiently explained by single-scattering processes. The formation
of shadows by macroscopic roughness in the visible range indicates
that the roughness size scale is much larger than the wavelength,
above thousands of microns to few millimeters, if the break in fractality
according to the thermal roughness-scale is real. The specular component
may indicate that carbonates are widespread and will likely be present
in the collected sample. Nightingale's low albedo, on the other hand,
could suggest fewer rock faces larger than OCAMS pixel-size scattering
back to the observer, thus decreasing photometric variability. Therefore,
Nightingale's roughness size scale may be much smaller than Bennu's
average.}{\footnotesize \par}

\subsubsection*{\emph{\scriptsize{}Acknowledgements. P. H. H. thanks Dr. Alice Bernard
for her support. This work was funded by the DIM-ACAV+ project (�le-de-France,
France) and is based upon work supported by NASA under Contract NNM10AA11C
issued through the New Frontiers Program. The authors thank all teams
and people that helped in the accomplishment of the OSIRIS-REx mission.
P. H. H. thanks all people concerned in developing and keeping the
Cython language, and Numpy and Scipy libraries for Python. OLA and
funding for the Canadian authors were provided by the Canadian Space
Agency. Images and kernels will be available via the Planetary Data
System (PDS) (https://sbn.psi.edu/pds/resource/orex/). Data are delivered
to the PDS according to the OSIRIS-REx Data Management Plan available
in the OSIRIS-REx PDS archive. DTMs will be available in the PDS 1
year after departure from the asteroid.}}

\section*{{\small{}References}}

\bibliographystyle{aasjournal}
\bibliography{orex}

\section*{{\small{}Appendix A}}

{\footnotesize{}As we are dealing with unresolved shadowed surfaces
that are getting expressed into a detector by a single pixel intensity
$I_{\lambda}(i,e,\alpha,\varphi)$, the pixel intensity can get split
into two terms (the meanings of all variables are listed in the Table
\ref{tab:variables}):}{\footnotesize \par}
\noindent \begin{center}
{\footnotesize{}
\[
I_{\lambda}(i,e,\alpha,\varphi)\varOmega_{T}=\sum_{j}\mu_{0}R_{a}^{(j)}(\alpha)D_{a}^{(j)}(i,e,\varphi)\varOmega_{e}^{(j)}S_{\lambda}+
\]
}
\par\end{center}{\footnotesize \par}

\noindent \begin{center}
{\footnotesize{}
\begin{equation}
+\sum_{j}I_{n}\left(1-\varOmega_{e}^{(j)}\right)
\end{equation}
}
\par\end{center}{\footnotesize \par}

{\footnotesize{}In the equation above we follow the assumption that
the reflected intensity can be decomposed in two functions: a scattering
phase function $R_{a}^{(j)}(\alpha)$ and photometric-topographic
``disk function'' $D_{a}^{(j)}(i,e,\varphi)$ \citep{2011P&SS...59.1326S}.
The first term represents the total intensity contribution of all
visible and illuminated facet area covered by the pixel \citep{2015P&SS..118..250W}.
The second term is the second-order scattered intensity contribution
of all of the surface not directly illuminated but yet visible. This
quantity is approximated to a diffusive reflectance \citep{Hapke2012}
by assuming that the surrounding illuminated surfaces can be treated
as an isotropic light source (reflected light from all surrounding
illuminated topography). By considering that all facet reflectances
$R_{a}(\alpha)$ are albedo-homogeneous, and that every $r_{0}$ is
isotropic and homogeneous, we can obtain the phase function reflectance
$R_{a}(\alpha)$ per pixel by re-arranging:
\[
R_{a}(\alpha)=\left[\frac{I_{\lambda}(i,e,\alpha,\varphi)}{S_{\lambda}}-r_{0}\left(1-\sum_{j}\varOmega_{e}^{(j)}\right)\right]\times
\]
}{\footnotesize \par}
\noindent \begin{center}
{\footnotesize{}
\begin{equation}
\times\sum_{j}\frac{\varOmega_{T}}{D_{a}^{(j)}(i,e,\varphi)\mu_{0}\varOmega_{e}^{(j)}}
\end{equation}
}
\par\end{center}{\footnotesize \par}

{\footnotesize{}}
\begin{table}
{\footnotesize{}\caption{{\footnotesize{}\label{tab:variables}Variable in equations of Appendix
A.}}
}{\footnotesize \par}
\begin{centering}
\begin{tabular}{c>{\centering}p{5cm}}
\hline 
{\footnotesize{}Variable} & {\footnotesize{}Description}\tabularnewline
\hline 
\hline 
{\footnotesize{}$j$} & {\footnotesize{}facet index}\tabularnewline
\hline 
{\footnotesize{}$I_{\lambda}$} & {\footnotesize{}Pixel Intensity}\tabularnewline
\hline 
{\footnotesize{}$I_{n}$} & {\footnotesize{}retro-scattered intensity from shadowed surface}\tabularnewline
\hline 
{\footnotesize{}$R_{a}$} & {\footnotesize{}phase function reflectance or the corrected radiance
factor ratio}\tabularnewline
\hline 
{\footnotesize{}$D_{a}$} & {\footnotesize{}``disk function'' ratio}\tabularnewline
\hline 
{\footnotesize{}$\varOmega_{T}$} & {\footnotesize{}Pixel solid angle}\tabularnewline
\hline 
{\footnotesize{}$\varOmega_{e}$} & {\footnotesize{}Fraction of Solid angle of a surface element}\tabularnewline
\hline 
{\footnotesize{}$\mu_{0}$} & {\footnotesize{}cosine of incidence angle}\tabularnewline
\hline 
{\footnotesize{}$S_{\lambda}$} & {\footnotesize{}Solar irradiance}\tabularnewline
\hline 
{\footnotesize{}$r_{0}$} & {\footnotesize{}diffusive reflectance}\tabularnewline
\hline 
\end{tabular}
\par\end{centering}
\end{table}
{\footnotesize{}Operations to remove the photometric-topographic effect,
the so-called ``disk function'', and the contribution of shadows
were calculated using the equation above. In our paper the $\mu_{0}D_{a}^{(j)}(i,e,\varphi)$
joint term is replaced by $L_{r}(i,e,\alpha,\varphi)$ calculated
from the semi-numerical roughness model (Section 4), therefore $R_{a}(\alpha)\longrightarrow R_{a}$,
the corrected RADF ratio. This operation reduces the dependence of
the radiance to any large-scale shadow that the DTM is capable of
tracing, leaving out only the intrinsic dependence on sub-facet roughness
and scattering.}{\footnotesize \par}

{\footnotesize{}The diffusive reflectance is somewhat related to the
second-order scattering albedo. During our image treatment, instead
of leaving it as another free parameter, we set $r_{0}$ very small
(=1e-5); therefore, macro-shadows will have a minimal contribution
to the pixel RADF. The only error that we may incur is to overestimate
the RADF for some heavily shadowed pixels.}{\footnotesize \par}
\end{document}